\documentclass[12pt]{article}

\usepackage{fullpage}
\usepackage{amssymb}
\usepackage{amsthm}
\usepackage{amsmath}
\usepackage{booktabs}
\usepackage{graphicx}
\usepackage{natbib}
\usepackage[multiple]{footmisc}
\usepackage[hyphens]{url}
\usepackage{color,soul}
\usepackage{multirow}
\usepackage[top=1in, bottom=1.05in, left=1in, right=1in]{geometry}
\usepackage[linesnumbered,ruled]{algorithm2e}
\usepackage{tikz}
\usepackage[labelformat=simple]{subcaption}

\usetikzlibrary{arrows,shapes,calc}

\tikzstyle{block} = [rectangle, draw, fill=white, 
    text width=7cm, text centered, rounded corners, minimum height=3em]

\newcommand{\bfx}[1]{{\bf #1}}

\title{Analysis of Prescription Drug Utilization with Beta Regression Models}

\author{Guojun Gan\thanks{Corresponding author, Department of Mathematics, University of Connecticut, 341 Mansfield Road, Storrs, CT, 06269-1009, USA. Email: \texttt{guojun.gan@uconn.edu}.} \and Emiliano A. Valdez\thanks{Department of Mathematics, University of Connecticut, 341 Mansfield Road, Storrs, CT, 06269-1009, USA. Email: \texttt{emiliano.valdez@uconn.edu}.}}

\begin{document}

\maketitle

\begin{abstract}
The healthcare sector in the U.S. is complex and is also a large sector that generates about 20\% of the country's gross domestic product. Healthcare analytics has been used by researchers and practitioners to better understand the industry. In this paper, we examine and demonstrate the use of Beta regression models to study the utilization of brand name drugs in the U.S. to understand the variability of brand name drug utilization across different areas. The models are fitted to public datasets obtained from the Medicare \& Medicaid Services and  the Internal Revenue Service. Integrated Nested Laplace Approximation (INLA) is used to perform the inference. The numerical results show that Beta regression models can fit the brand name drug claim rates well and including spatial dependence improves the performance of the Beta regression models. Such models can be used to reflect the effect of prescription drug utilization when updating an insured's health risk in a risk scoring model.

\bigskip
\noindent \textbf{Keywords}: Beta regression, healthcare analytics, random effects, spatial modeling, 

\end{abstract}

\section{Introduction}

In the pharmaceutical industry, a brand name drug is a prescription drug that is developed and patented by a company. The patent gives  the company a period of market exclusivity, during which the company sells the brand name drug at a significantly high price. For more information about drug pricing, readers are referred to \cite{schoonveld2015}. When the patent expires, generic versions of the brand name drug are marketed at lower prices by other companies. Generally, most states allow pharmacists to substitute brand name drugs with generic versions, unless otherwise directed by physicians.   Nevertheless, according to the Association for Accessible Medicines (AAM) \citep{klinger2018drug},  the generics account for 89\% of prescriptions dispensed but only 26\% of total drug costs in the U.S. Recent annual prescription spending in the United States has been on the rise. According to the NHE (National Health Expenditure) fact sheet \footnote{\url{https://www.cms.gov/Research-Statistics-Data-and-Systems/Statistics-Trends-and-Reports/NationalHealthExpendData/NHE-Fact-Sheet}. Accessed on July 6, 2020.},  prescription drug spending in the U.S. increased 2.5\% to \$335.0 billion in 2018, faster than the 1.4\% growth in 2017.

This large disparity in the costs between brand name and generic drugs motivated us to investigate for the presence of feature variables, e.g., location, income, demographics, that drive the decision to choose one type over the other. It is also our intent to prescribe a model that health insurers can replicate to reflect these decision drivers into a risk adjustment program  \citep{duncan2011}, a mechanism widely popular for assessing the health risk of an insured \citep{hileman2016score}. This mechanism is an umbrella of commercially designed risk assessment models that are sometimes called risk scoring models. While models do vary, they share a common objective of predicting the health care cost per member per year using prior year's medical utilization and expenditures as well as additional risk factors such as location and demographic.

The decision to prescribe/purchase generic or brand name drugs has been studied before. \cite{brinberg1984drug} used two behavioral intension models to examine the decision to purchase generic prescription drugs and found several differences between individuals who intent to purchase generic prescription drugs with those who do not. For example, nonintenders believed it was less likely that generic drugs were a safe product than intenders. \cite{hamann2013drug} studied psychiatrists' decision making between generic and branded antipsychotics and found that psychiatrists were more likely to choose branded drugs when imagining choosing the drug for themselves. \cite{hassali2014drug} reviewed experiences of implementing generic medicine policy in eight countries, including the U.S. The review indicates that pharmacists play an essential role in promoting generic medicines as they substituted 83.8\% of prescriptions that allowed substitution and that generic prescribing is still not common practice in the U.S. because many physicians have negative perceptions about generic medicines and lack in-depth knowledge about the bioequivalence concept
applied in the U.S. \cite{kesselheim2016drug} studied variations in patients' perceptions and use of generic drugs based on a national survey and found a substantial shift that more patients have positive views of generic drugs.

In the aforementioned work, \cite{brinberg1984drug}, \cite{hamann2013drug}, and \cite{kesselheim2016drug} used survey data to study the behavior of the patients and physicians. The review conducted by \cite{hassali2014drug} was based on literature search using several electronic databases and search engines such as ISI ( Institute for Scientific Information) Web of Knowledge and Google. Public pharmaceutical drug utilization data have not been used to study the factors driving the decision of choosing one type over the other.

In this paper, we investigate the rates of brand name drug claims in the U.S. In particular, we investigate the variations of the brand name drug claims rates in different areas of the U.S. Our study supplements the aforementioned studies based on survey data. Since rates are values in the interval $(0,1)$, the Beta regression model \citep{ferrari2004beta} is most suitable where covariates can be introduced to account for heterogeneity.

The remaining part of the paper is organized as follows. In Section \ref{sec:review}, we review some academic work related to health insurance. In Section \ref{sec:data}, we describe the data used in our study. In particular, we describe how the data is processed before models are fitted. In Section \ref{sec:model}, we present four Beta regression models for fitting the data. In Section \ref{sec:result}, we provide numerical results of the proposed Beta regression models. In Section \ref{sec:con}, we conclude the paper with some remarks.

\section{Related work}\label{sec:review}

Healthcare is an important practicing area of actuaries due to the existence of health insurance. In this section, we review some work that is related to health insurance analytics.

Early work includes \cite{bolnick2004health}, \cite{petertill2005health}, and \cite{bachler2006health}. \cite{bolnick2004health} proposed a framework for long-term healthcare cost projections by incorporating key healthcare cost drivers such as life expectancy, biological morbidity, and economic morbidity. Using the framework, \cite{bolnick2004health} considered various plausible future scenarios that encompass a reasonable range of healthcare cost outcomes. \cite{petertill2005health} studied the relative significance of aging as a driver of healthcare cost beyond age fifty and used Medicare data to draw general conclusions that utilization and cost differ by age.
\cite{bachler2006health} studied the impact of chronic and nonchronic insured members on cost trends and found that classification of chronic members can affect the trends significantly.  

Recently, there has been a lot more studies on healthcare analytics than ten years ago. \cite{duncan2011} devoted a book to healthcare analytics for actuaries. In particular, this book covers the essentials of health risk and case studies on the use of predictive modeling in risk adjustment. For example, Chapter 14 of this book discusses the risk adjustment method used by the Centers for Medicare and Medicaid Services (CMS). \cite{frees2011health} extended the standard two-part model to predict the frequency and amount of healthcare expenditures and used the data from the Medical Expenditure Panel Survey (MEPS) to calibrate and test the model.  \cite{shi2013health} built a game-theoretic model based on copulas to study the effect of managed care on healthcare utilization compared to traditional fee-for-service plans in private health insurance market. 

\cite{getzen2016health} developed a method to evaluate projections of future medical spending used by the U.S. Medicare and Medicaid programs and found that the recent set of projections (1998--2010) is more accurate than the older set of  projections (1986--1995) because the recent projections incorporate lagged macroeconomic effects. \cite{duncan2016health} investigated several statistical techniques (e.g., Lasso GLM, multivariate adaptive regression splines, random forests, decision trees, and boosted trees) for modeling future healthcare costs and found that the traditional regression approach does not perform well. 

\cite{huang2017health} compared different models and model selection methods for health costs in Episode Treatment Groups (ETGs) and found that random forest feature selection is preferable in terms of efficiency and accuracy. \cite{richardson2018health} proposed a Bayesian nonparametric regression model for predicting healthcare claims. Their numerical results show that the Bayesian nonparametric model outperforms standard linear regression and generalized Beta regression in terms of predictive accuracy. \cite{brockett2018health} investigated the use of Data Envelopment Analysis (DEA), which is a method developed in management science to measure relative efficiency of organizations, to assess the potential savings of Medicare. Their analysis shows that Medicare Advantage plans are more efficient in reducing health expenditures than the original Medicare. In \cite{brockett2019health}, the authors used linear regression models to estimate the effect of medical loss ratio (MLR) and efficiency on the quality of care for the Medicaid managed care plans. The results show that the effect of medical services efficiency on the quality of care is insignificant. \cite{kan2019health} explored the use of machine learning techniques for risk adjustment and found that penalized regression performs better than ordinary least squares in predicting healthcare costs. 

Our work presented in this paper is relative to healthcare analytics but is different from the work mentioned above. In particular, our work examined Beta regression model with spatial dependence to model the percentages of bran name drug claims. The methods and findings of this paper can be used by practitioners in their risk modeling and adjustment.

\section{Description of the Data}\label{sec:data}

We use a public dataset called the Part D Prescriber Public Use File (PUF)  from the Centers for Medicare \& Medicaid Services (CMS) and another public dataset called Individual Income Tax Statistics from the IRS (Internal Revenue Service). 

The data in the Part D Prescriber PUF cover calendar years 2013 through 2016 and contain information on prescription drug events (PDEs) incurred by Medicare beneficiaries with a Part D prescription drug plan. The data consist of a detailed data file and two summary tables. The ``Part D Prescriber Summary Table'' contains overall drug utilization, drug costs, and beneficiary counts organized by National Provider Identifier (NPI). In this paper, we use the 2016 Part D Prescriber Summary Table, which contains 1,131,550 records and 84 variables \footnote{The file name is \texttt{PartD\_Prescriber\_PUF\_NPI\_16.txt} and it is available from \url{https://www.cms.gov/Research-Statistics-Data-and-Systems/Statistics-Trends-and-Reports/Medicare-Provider-Charge-Data/PartD2016}. Accessed on Jan 20, 2020.}.  The Part D Prescriber Summary table contains information about the brand and generic drug claims at the provider level. Other information includes the zip code of providers, the count of beneficiaries, the average age of beneficiaries, and the average risk score of beneficiaries.

In this paper, our interest is on studying variations of the brand name drug claims rates in different areas of the U.S. To that end, we need to aggregate the data to different areas. However, the data can be aggregated to different levels: the state level, the zip code level, or some customized level between the state level and the zip code level. On the one hand, aggregating the data to the state level is not desirable for the following reasons:
\begin{itemize}
	\item The brand name drug claim rate exhibits variations within individual states. 
	\item The U.S. has only 50 states and aggregating the data to the state level produces only 50 data points.
\end{itemize}
On the other hand, aggregating the data to the zip code level is also not desirable:
\begin{itemize}
	\item The dataset contains about 20,000 different zip codes. Aggregating the data to the zip code level will produce about 20,000 data points. This will cause challenges in modeling the spatial effects as a large number of sites requires lots of parameters.
	\item Some zip codes do not correspond to geographical areas but large volume customers or post office boxes.
	\item Aggregating the data to the zip level produces highly volatile brand name drug claims rates. That is, the rates are highly volatile between zip codes.
\end{itemize}
As a result, we decide to aggregate the data to some customized level between the state level and the zip code level. 

\begin{figure}[htbp]
	\centering
	\includegraphics[width=0.95\textwidth]{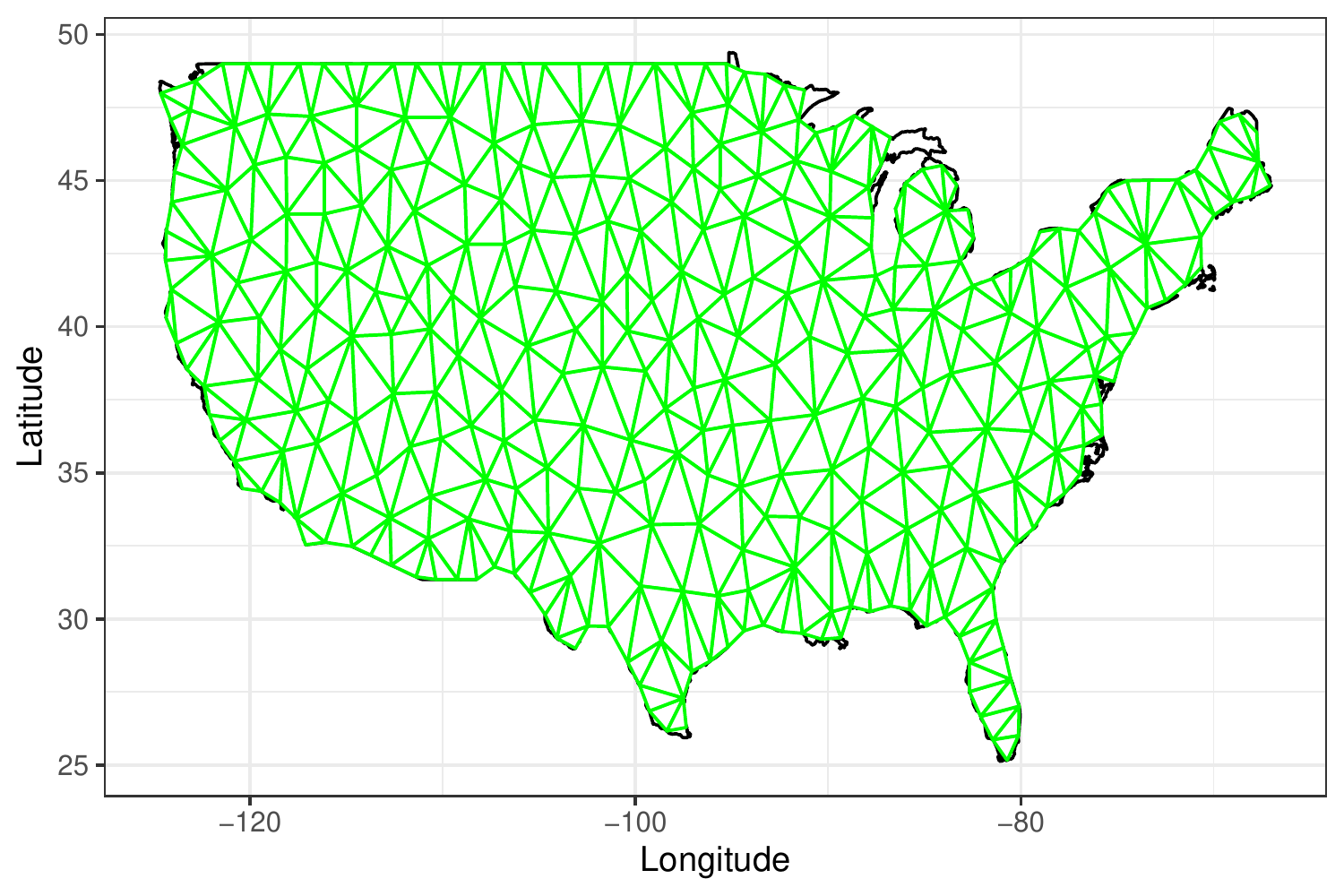}
	\caption{A mesh of the 48 contiguous states with 530 triangles.}\label{fig:mesh}
\end{figure}

To aggregate the data to a customized level, we first need to split the U.S. land into small areas, i.e., create a mesh of the U.S. land. This can be done conveniently by using the R function \texttt{inla.mesh.2d} from the \texttt{INLA} package. Figure~\ref{fig:mesh} shows a mesh of the 48 contiguous states of the U.S. that consists of 530 triangles. Aggregating the data to the 530 triangular areas will produce a dataset with about 530 data points. This is computationally reasonable for modeling spatial effects.

Aggregating the data to the triangles shown in Figure~\ref{fig:mesh} is done as follows. First, we obtain the longitudes and the latitudes of zip codes from the R package \texttt{noncensus}, which contains regional information and demographic data determined by the U.S. Census Bureau. Note that each zip code is associated with a longitude and a latitude. Second, we determine the triangles to which the zip codes belong by using the longitudes and the latitudes. Finally, we aggregate the data based on the indices of the triangles. 

The Part D Summary Table contains average ages and average risk scores of the beneficiaries. The average risk scores are average HCC (Hierarchical Condition Category) risk scores, which estimate how beneficiaries' FFS (Fee-For-Service) spending will compare to the overall average of the entire Medicare population.  Before aggregating the data, we convert average ages and average risk scores to total ages and total risk scores by multiplying the average numbers with the count of beneficiaries. After aggregation is done, we convert total ages and total risk scores to average ages and average risk scores by dividing the total number by the aggregated count of beneficiaries. The Part D Summary Table also contains missing values. We remove missing values before aggregating the data.

\begin{figure}[htbp]
	\centering
	\begin{subfigure}[t]{0.49\textwidth}
		\includegraphics[width=\textwidth]{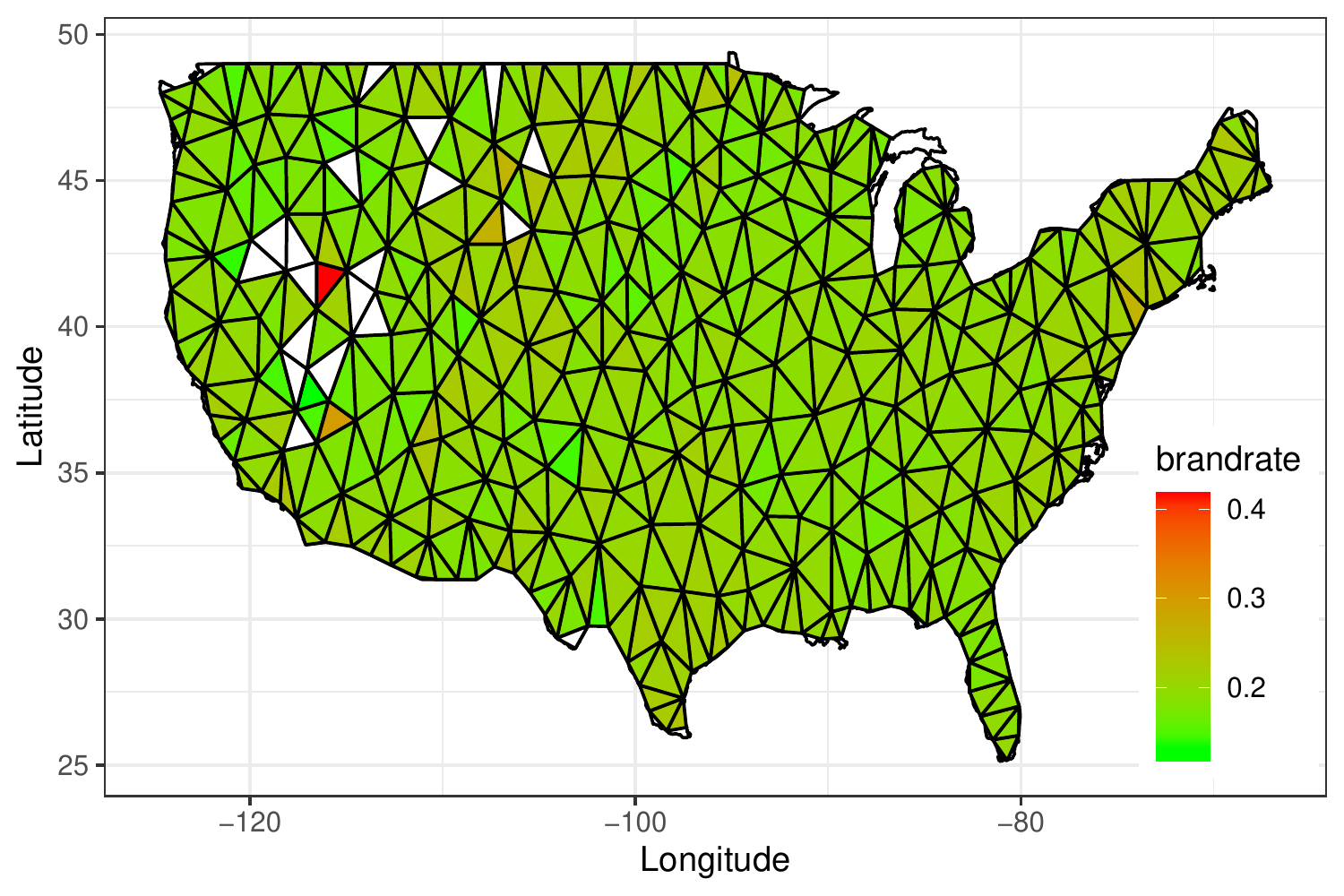}
		\caption{}\label{fig:rate}
	\end{subfigure} \hfill
	\begin{subfigure}[t]{0.49\textwidth}
		\includegraphics[width=\textwidth]{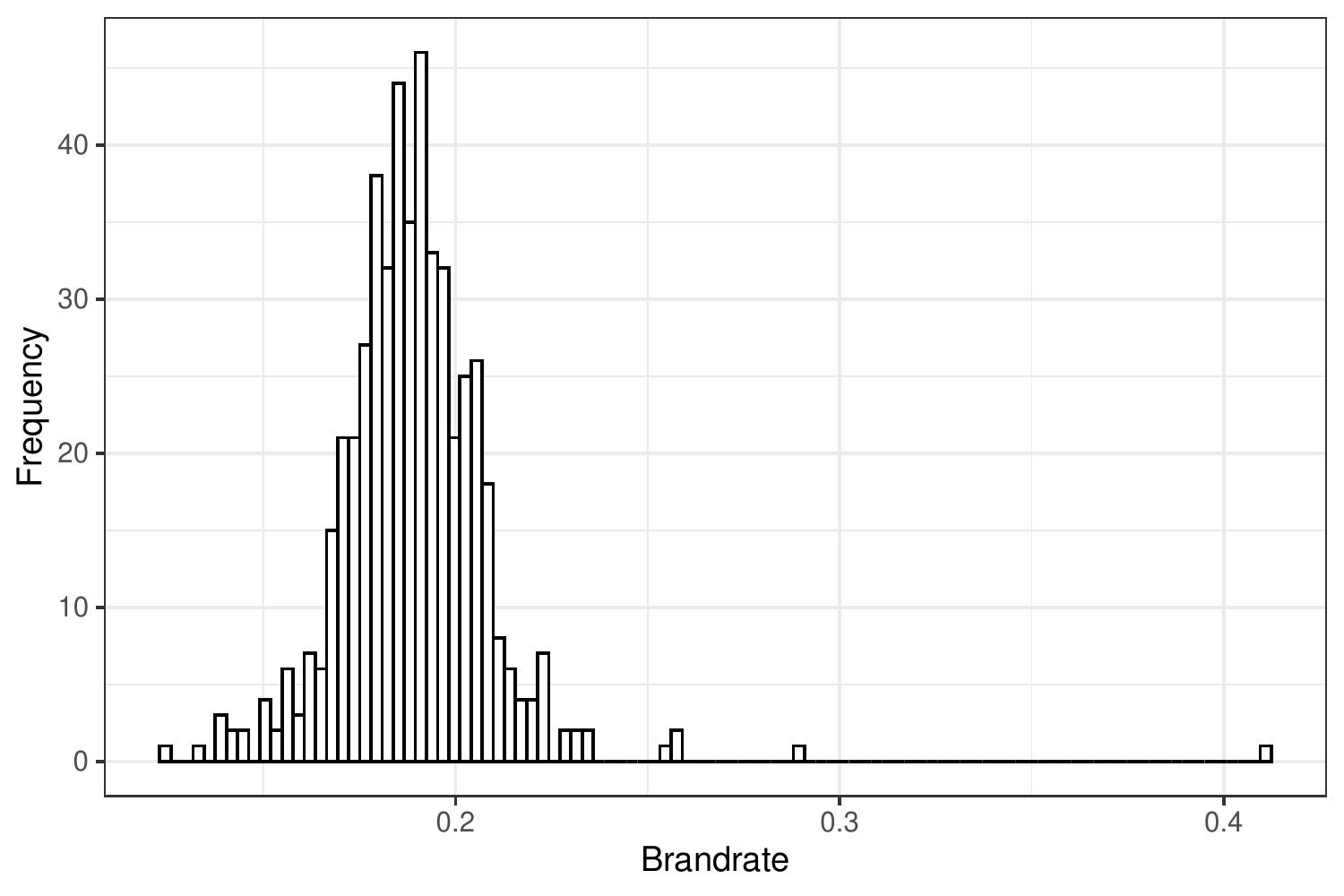}
		\caption{}\label{fig:ratehist}
	\end{subfigure}
\caption{Distributions of the brand name drug claim rates. White triangles mean that data are not available in these areas.}\label{fig:ratedist}
\end{figure}

Figure \ref{fig:rate} shows the distribution of the brand name drug claim rates in different triangular areas. From the figure, we see that 19 triangular areas do not have the data. As a result, the aggregated dataset contains 511 observations. From the figure, we also see that some triangles in the middle are light red. This suggests that modeling the spatial effects may improve the fitting of Beta regression models to the data.

\begin{table}[htbp]
	\centering
	\caption{Summary statistics of the Part D data.}\label{tbl:data1}
	\begin{tabular}{lrrrrrr}
		\toprule
		& \textbf{Min} & \textbf{1st Q} & \textbf{Median} & \textbf{Mean} & \textbf{3rd Q} & \textbf{Max} \\
		\midrule
		\texttt{brandrate} &      0.1253  &      0.1784  &      0.1886  &      0.1890  &      0.1986  &      0.4118  \\
		\texttt{avgage} &         65.79  &         70.06  &         70.89  &         70.98  &         71.87  &         80.00  \\
		\texttt{avgscore} &      0.9129  &      1.2102  &      1.3240  &      1.3170  &      1.4199  &      1.8166  \\
		\bottomrule
	\end{tabular}%
\end{table}%

Table \ref{tbl:data1} shows the summary statistics of the aggregated data. From the table, we see that the brand name drug claim rate varies from 12.53\% to 41.18\% among the 511 areas. The range of the average age of beneficiaries is from 65.79 to 80. The average risk score has a range of 0.9129 to 1.8166. The median and mean values indicate that these variables are pretty symmetrical.

Another data source we use in our study is the 2016 individual income tax statistics by zip codes \footnote{The file name is \texttt{16zpallagi.csv} and it is available from \url{https://www.irs.gov/statistics/soi-tax-stats-individual-income-tax-statistics-2016-zip-code-data-soi}. Accessed on Jan 12, 2020.}. This dataset contains 29,874 records, each of which is described by 144 variables. It contains information about the number of returns and the amount of returns in different categories. For example, it contains the number of returns with unemployment compensation and the unemployment compensation amount. The return counts and amounts are aggregated to the zip code level. Although this datast does not contain all information about an area, it does provide some demographic and economic information, which might be useful for explaining the variation of the brand name drug claim rate. For example, the number of returns from an area is related to the population of the area. 

We aggregate the tax data to the 530 triangular areas in the same way as we aggregate the Part D data. After aggregating the data, we divide the total dollar amount of an area by the corresponding number of returns to get the average dollar amount of the area. The total number of returns and the average amounts in different categories are used in the data analysis. Since this dataset has 144 variables, we do not show the summary statistics of these variables here. 

\section{Models}\label{sec:model}

In this section, we describe some Beta regression models. In particular, we present four models: the basic Beta regression model, the Beta regression model with random effects, the Beta-Besag model, and the Beta-BYM model. 

The density function of the Beta distribution is typically defined as \citep{klugman2012}:
\begin{equation}\label{eq:beta}
f(y; p, q) = \dfrac{\Gamma(p+q)}{\Gamma(p)\Gamma(q)}y^{p-1}(1-y)^{q-1},\quad 0<y<1,
\end{equation}
where $p>0$ and $q>0$ are shape parameters. The Beta distribution defined in Equation \eqref{eq:beta} has been reparameterized by using its mean and dispersion as parameters. 
\begin{equation}\label{eq:beta2}
f(y;\mu,\phi)=\dfrac{\Gamma(\phi)}{\Gamma(\mu \phi)\Gamma((1-\mu) \phi)}y^{\mu \phi-1}(1-y)^{(1-\mu) \phi -1}, \quad 0<y<1,
\end{equation}
where $0<\mu<1$ and $\phi>0$. The shape parameters can be obtained from the mean and the dispersion as follows: $p=\phi\mu$ and $q=\phi(1-\mu)$.

The basic Beta regression model is described as follows. Suppose that we have $n$ observations. For $i=1,2,\ldots,n$, let $\bfx{x}_i=(x_{i1}, x_{i2}, \ldots,x_{ik})^T$ and $y_i$ be the vector of $k$ regressors and the response in the $i$th case, respectively. The responses $y_1,y_2,\ldots, y_n$ are assumed to form a random sample such as
\[
y_i\sim Beta(\mu_i, \phi).
\]
The mean $\mu_i$ is linked to the regressors as follows:
\begin{equation}\label{eq:mui}
g(\mu_i) = \eta_i = \bfx{x}_i^T \boldsymbol{\beta},
\end{equation}
where $g(\cdot)$ is the link function and $\boldsymbol{\beta}=(\beta_1,\beta_2,\ldots,\beta_k)^T$ is the vector of regression coefficients. The variance of $y_i$ can be estimated as
\begin{equation}\label{eq:varyi}
\operatorname{Var}(y_i) = \dfrac{\mu_i(1-\mu_i)}{1+\phi}.
\end{equation}

Random effects models are commonly used to analyze areal or spatial data. For example, \cite{tufvesson2019spatial} applied random effects models to model car insurance data with geographical locations of the policyholders. There are two types of random effects models: unstructured and structured. Unstructured random effects models assume independence of the random effects, while structured random effects models allow for spatial dependence. 

The second model we consider is a Beta regression model with a single random effect. In this model, a random effect is introduced to the linear predictor as follows:
\begin{equation}\label{eq:mui2}
g(\mu_i) = \eta_i = \bfx{x}_i^T \boldsymbol{\beta} + v_i,
\end{equation} 
where $v_i$ are i.i.d Gaussian noise, i.e.,
\begin{equation}\label{eq:vi}
v_i \sim N\left(0, \dfrac{1}{\psi_1}\right),
\end{equation}
where $\psi_1$ is a precision parameter. This model can help to control for unobserved heterogeneity  when the heterogeneity is not correlated with independent variables. The basic Beta regression model assumes that the heterogeneity is correlated with independent variables.

\begin{figure}[htbp]
	\centering
	\includegraphics[width=0.95\textwidth]{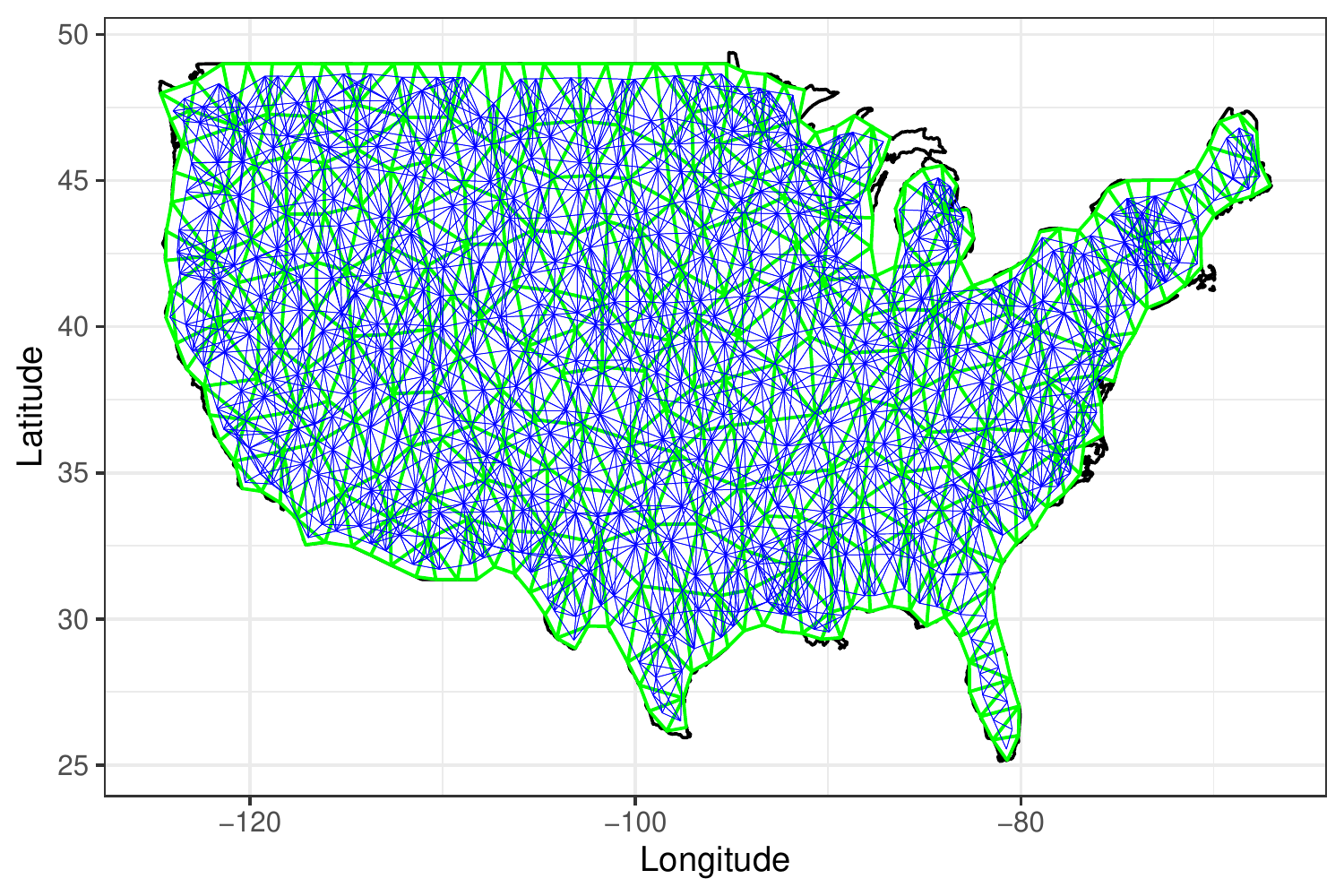}
	\caption{The neighborhood graph used to model spatial effects.}\label{fig:madj}
\end{figure}

The third model we consider is the Besag model, which introduces a structured random effect to the linear predictor:
\begin{equation}\label{eq:mui3}
g(\mu_i) = \eta_i = \bfx{x}_i^T \boldsymbol{\beta} + u_i,
\end{equation}
where $u_i$s follow a CAR(1) model, i.e., an order-1 conditional autoregressive model \citep{besag1995car}. Building a CAR model requires a neighborhood graph, which tells which areas are neighbors to each other. Figure~\ref{fig:madj} shows a neighborhood graph where areas that are neighbors are connected by blue lines. In our models, we assume that triangles that share a vertex or a side are neighbors to each other. 

Given a neighborhood graph, the CAR(1) model assumes that the random effect in an area is related to the random effects in the area's neighbors through a conditional Gaussian distribution:
\begin{equation}\label{eq:car1}
u_i\vert \{u_j: j\in A_i\} \sim N\left( \dfrac{1}{\vert A_i\vert} \sum_{j\in A_i} u_j,  \dfrac{1}{\psi_2 \vert A_j\vert}\right),
\end{equation}
where $A_i$ is the set of indices of neighbors of the $i$th area, $\vert A_i\vert$ denotes the number of elements in $A_i$, and $\psi_2$ is a precision parameter. Note that the CAR(1) model can be written as a multivariate Gaussian model for $\mathbf{u}=(u_1,u_2,\ldots,u_n)$ \citep{besag1995car,wang2018}:
\begin{equation}
\mathbf{u} \sim N\left(\mathbf{0}, \dfrac{1}{\psi_2}\mathbf{Q}^{-1}\right),
\end{equation}
where $\mathbf{Q}$ is a highly sparse matrix defined by
\[
Q_{ij} = \begin{cases}
\vert A_i\vert , & \hbox{if } i=j,\\
-1, & \hbox{if } j \in A_i,\\
0, & \hbox{otherwise.}
\end{cases}
\]

The fourth model is the BYM (Besag-York-Molli\'{e}) model, which combines an unstructured random effect and a structured random effect \citep{besag1991bym}:
\begin{equation}\label{eq:mui4}
g(\mu_i) = \eta_i = \bfx{x}_i^T \boldsymbol{\beta} + u_i + v_i,
\end{equation}
where $u_i$s are modeled by a CAR(1) model as specified in Equation \eqref{eq:car1} and $v_i$s are i.i.d Gaussian noise as specified in Equation \eqref{eq:vi}. The structured effects $\mathbf{u}$ model the spatial effect. The unstructured effects $\mathbf{v}=(v_1,v_2,\ldots,v_n)$ are used to model additional random variations that are not explained by geographical locations. Between the two effects $\textbf{u}$ and $\textbf{v}$,  a larger effect of $\mathbf{v}$ in the model means that less exchange of information between areas is allowed. The relative strengths of the unstructured effects $\mathbf{v}$ and the structured effects $\mathbf{u}$ are controlled by the two hyperparameters $\psi_1$ and $\psi_2$, respectively.

\begin{table}[htbp]
	\centering
	\caption{Linear predictors of the four models.}\label{tbl:models}
	\begin{tabular}{ll}
		\toprule
		\textbf{Model name} & \textbf{Linear predictor} \\
		\midrule
		BetaReg & Equation \eqref{eq:mui} \\
		BetaRE & Equation \eqref{eq:mui2} \\
		BetaBesag & Equation \eqref{eq:mui3} \\
		BetaBYM & Equation \eqref{eq:mui4} \\
		\bottomrule
	\end{tabular}
\end{table}

Table \ref{tbl:models} summarizes the four Beta regression models described above. The first two models do not consider spatial dependence, while the last two models allow for spatial dependence.

\begin{figure}[htbp]
	\centering
	\includegraphics[width=0.95\textwidth]{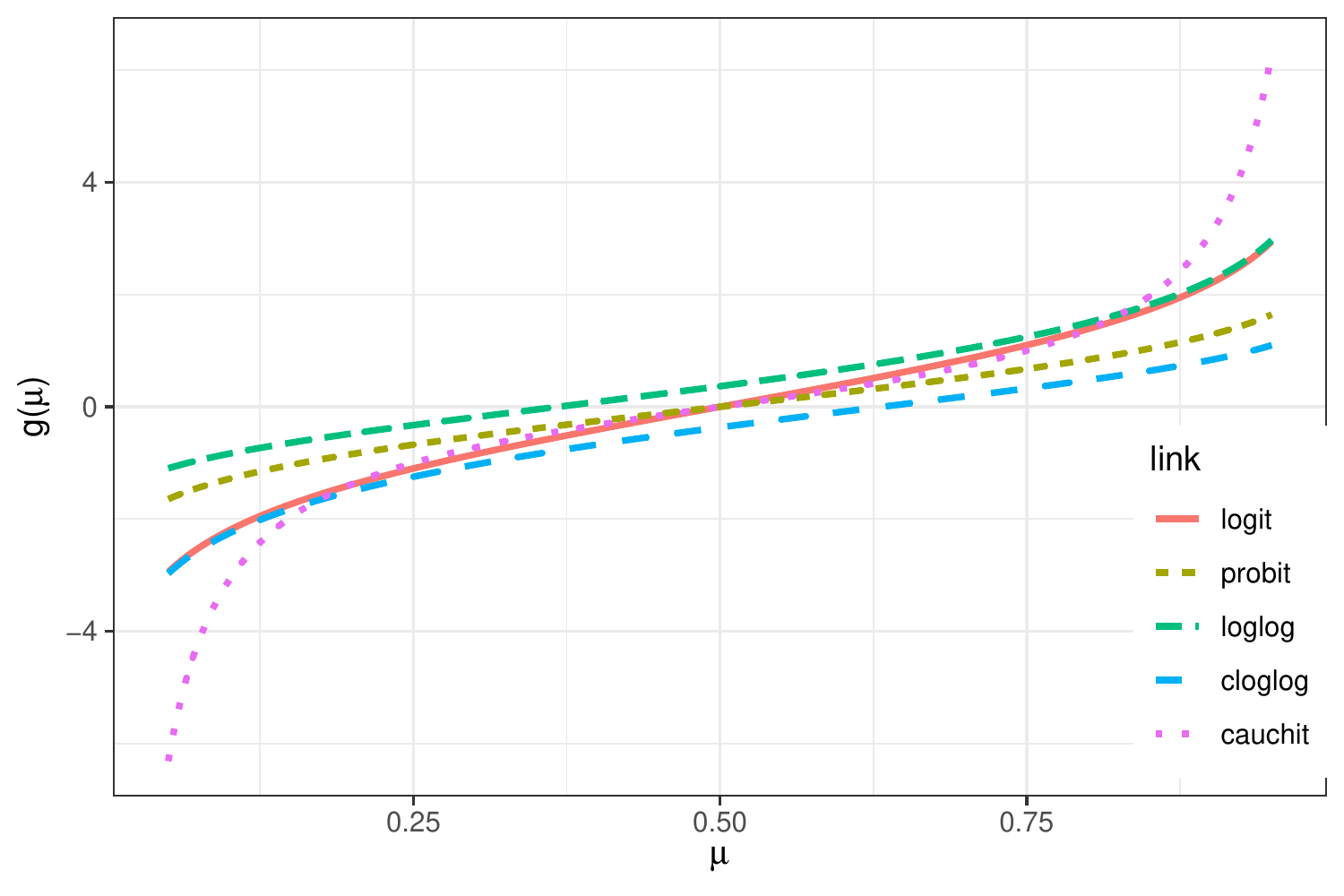}
	\caption{Some link functions for Beta regression models.}\label{fig:link}
\end{figure}

There are several choices of the link function \citep{ferrari2004beta}:
\begin{itemize}
	\item the logit link: $g(\mu)=\log \dfrac{\mu}{1-\mu}$;
	\item the probit link: $g(\mu) = \Phi^{-1}(\mu)$, where $\Phi(\cdot)$ is the standard normal distribution function;
	\item the log-log link: $g(\mu) = -\log(-\log(\mu))$;
	\item the complimentary log-log link: $g(\mu) = \log(-\log(1-\mu))$;
	\item the Cauchy link: $g(\mu)=\tan(\pi(\mu-0.5))$.
\end{itemize}
Figure \ref{fig:link} shows the plots of these link functions. The logit, probit, and Cauhy link function are symmetric around the point $(0.5, 0)$. The other two link functions are not symmetric around the point. 

\section{Results}\label{sec:result}

In this section, we present the results of fitting the four Beta regression models (see Table \ref{tbl:models}) to the brand drug claim data.

\subsection{Data transformation}

Before we fit the models, we transform the covariates by using the Yeo–Johnson transformation introduced by \cite{yeo2000pwoer}. The purpose of transforming the covariates is to reduce their magnitudes for the convenience of modeling fitting. The Yeo–Johnson transformation is similar to the power transformation but can handle zero and negative values. In this transformation, a value $x$ is transformed as follows:
\begin{equation}\label{eq:yeo}
x^\ast = \begin{cases}
\log(1+x), & \hbox{if } x\ge 0,\\
-\log(1-x), & \hbox{if } x<0.
\end{cases}
\end{equation}
We transform all covariates except for the average risk score before fitting the models.

\begin{figure}[htbp]
	\centering
	\begin{subfigure}{0.49\textwidth}
		\includegraphics[width=\textwidth]{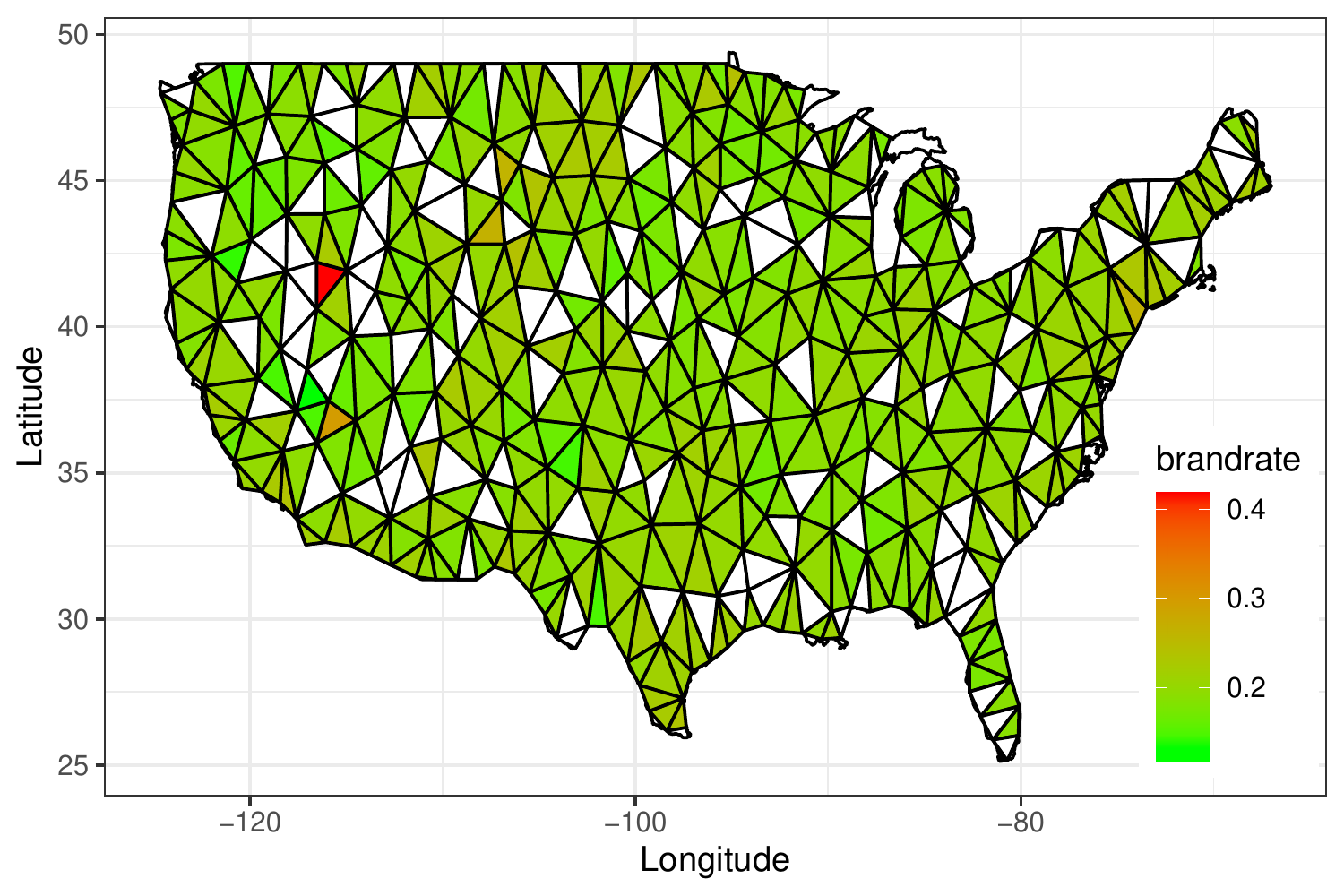}
		\caption{The training set.}
	\end{subfigure}
	\begin{subfigure}{0.49\textwidth}
		\includegraphics[width=\textwidth]{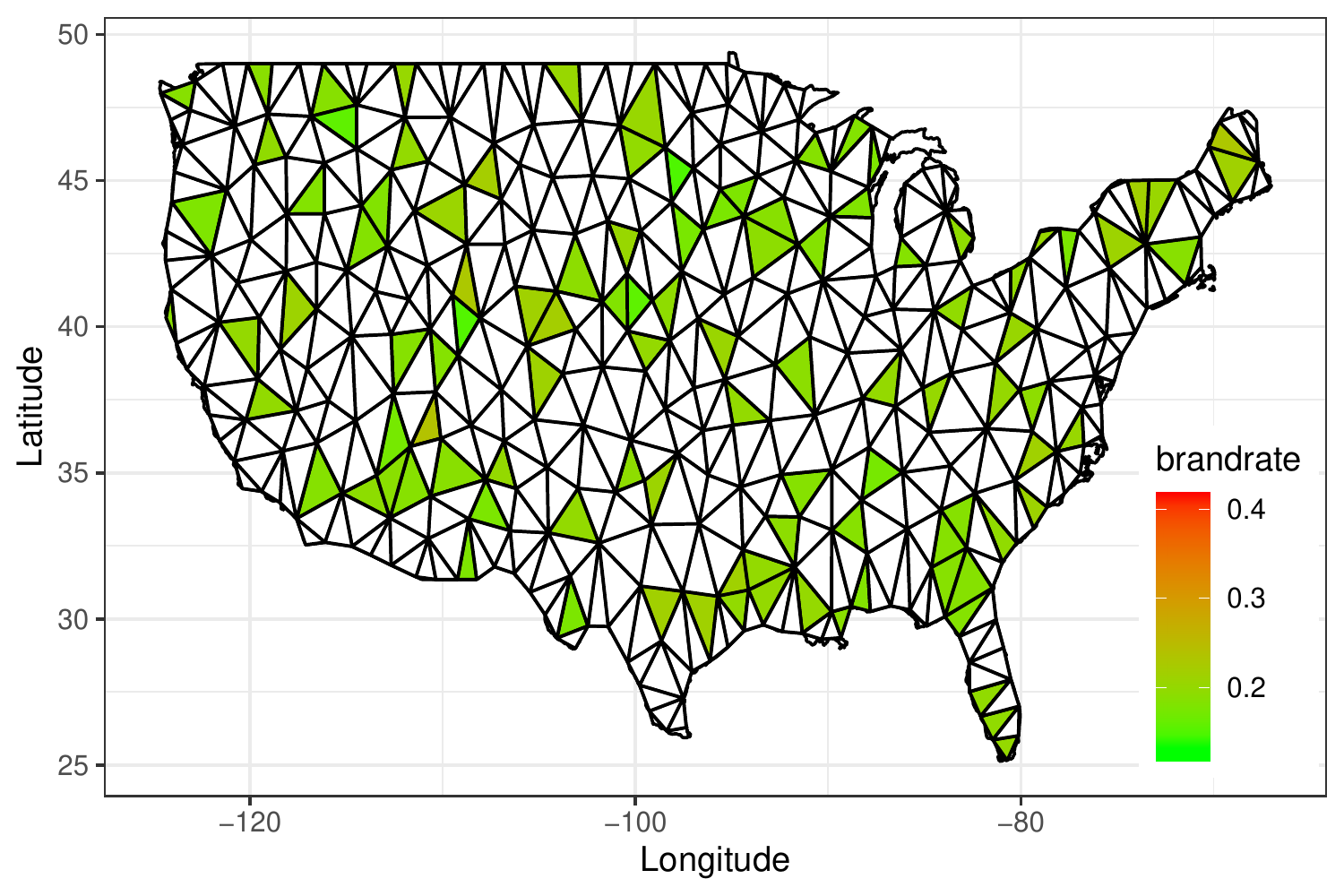}
		\caption{The test set.}
	\end{subfigure}
\caption{The distribution of the brand name drug claim rates for the training and test sets.}\label{fig:rate12}
\end{figure}

After transforming the data, we split the dataset into two sets: one for fitting the models and one for validating the models. We use 80\% of the data for fitting and the remaining 20\% for evaluation. Figure~\ref{fig:rate12} shows the distribution of the brand name drug claim rates for the training set and the test set.

\subsection{Selection of covariates}

The data contain 145 covariates, which include 143 from the individual tax income file and 2 (i.e., \texttt{avgage} and \texttt{avgscore}) from the Part D Summary Table. Including all these covariates in the models causes the parameter identification problem because many of these covariates are highly correlated. 

To select covariates for fitting the models, we use the lasso regularized generalized linear models (GLMs), which have been implemented in the R package \texttt{glmnet} \citep{friedman2010glmnet}. \cite{tufvesson2019spatial} used this method to select covariates. However, the package \texttt{glmnet} does not support Beta regression models. To circumvent this problem, we first create a new binary response variable by comparing the brand name drug claim rates to the average rate. If a rate is above the average, then the corresponding response value is 1. Otherwise, the response value is 0.  Then use lasso regularized logistic regression that is supported by \texttt{glmnet} to select covariates. We assume that the covariates that help to separate high and lower rates can also serve as useful predictors. 

\begin{figure}[htb]
	\centering
	\includegraphics[width=0.95\textwidth]{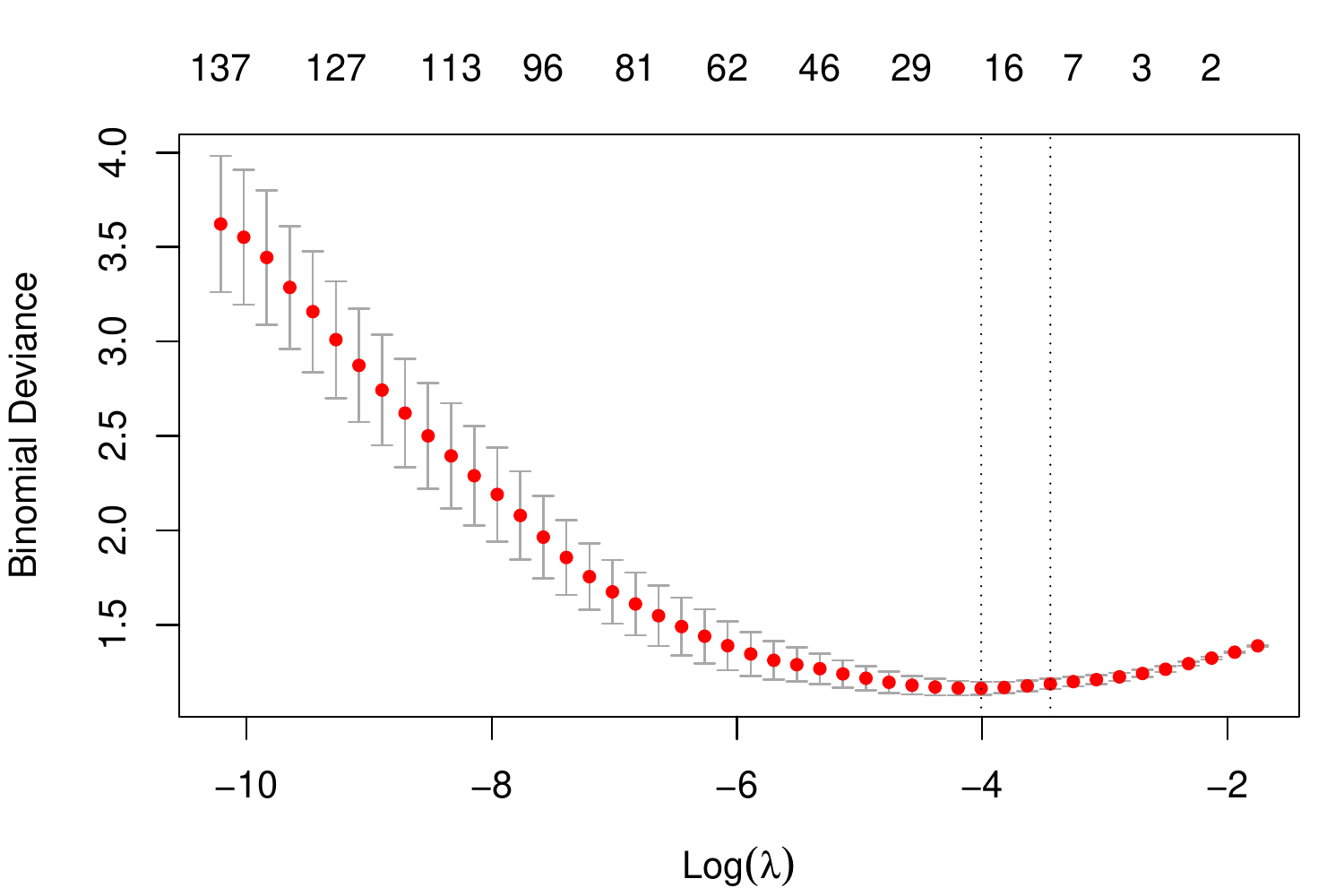}
	\caption{Results from a cross-valuation run of the lasso regularized logistic GLM based on the training set.}\label{fig:cvfit}
\end{figure}

Figure \ref{fig:cvfit} shows the results from a cross-validation run of the lasso regularized logistic GLM with different penalties $\lambda$. In the figure, the left dashed vertical line corresponds to the penalty $\lambda_{min}$ that minimizes the cross-validation error. The second dashed vertical line corresponds to the smallest penalty $\lambda_{1se}$ with the error that is within one standard deviation of the minimal error. We use the model with the penalty $\lambda_{1se}$ to select covariates. This gives us 12 covariates, which are described in Table \ref{tbl:covariate}. These 12 selected covariates are used to fit the Beta regression models.

\begin{table}[htbp]
	\centering
	\caption{Descripton of the selected covariates.}\label{tbl:covariate}
	\begin{tabular}{ll}
		\toprule
		\textbf{Covariate} & \textbf{Description} \\
		\midrule
		\texttt{avgscore} & Average risk score \\
		\texttt{VITA} & Number of volunteer income tax assistance (VITA) prepared returns  \\
		\texttt{A00900} & Business or professional net income (less loss) amount \\
		\texttt{A02300} & Unemployment compensation amount  \\
		\texttt{A03150} & Individual retirement arrangement payments amount \\
		\texttt{A03230} & Tuition and fees deduction amount \\
		\texttt{A18450} & State and local general sales tax amount \\
		\texttt{A18800} & Personal property taxes amount \\
		\texttt{A07230} & Nonrefundable education credit amount \\
		\texttt{A85770} & Total premium tax credit amount \\
		\texttt{A11070} & Additional child tax credit amount \\
		\texttt{A11902} & Overpayments refunded amount  \\
		\bottomrule
	\end{tabular}%
\end{table}%

\begin{figure}[htbp]
	\centering
	\includegraphics[width=0.8\textwidth]{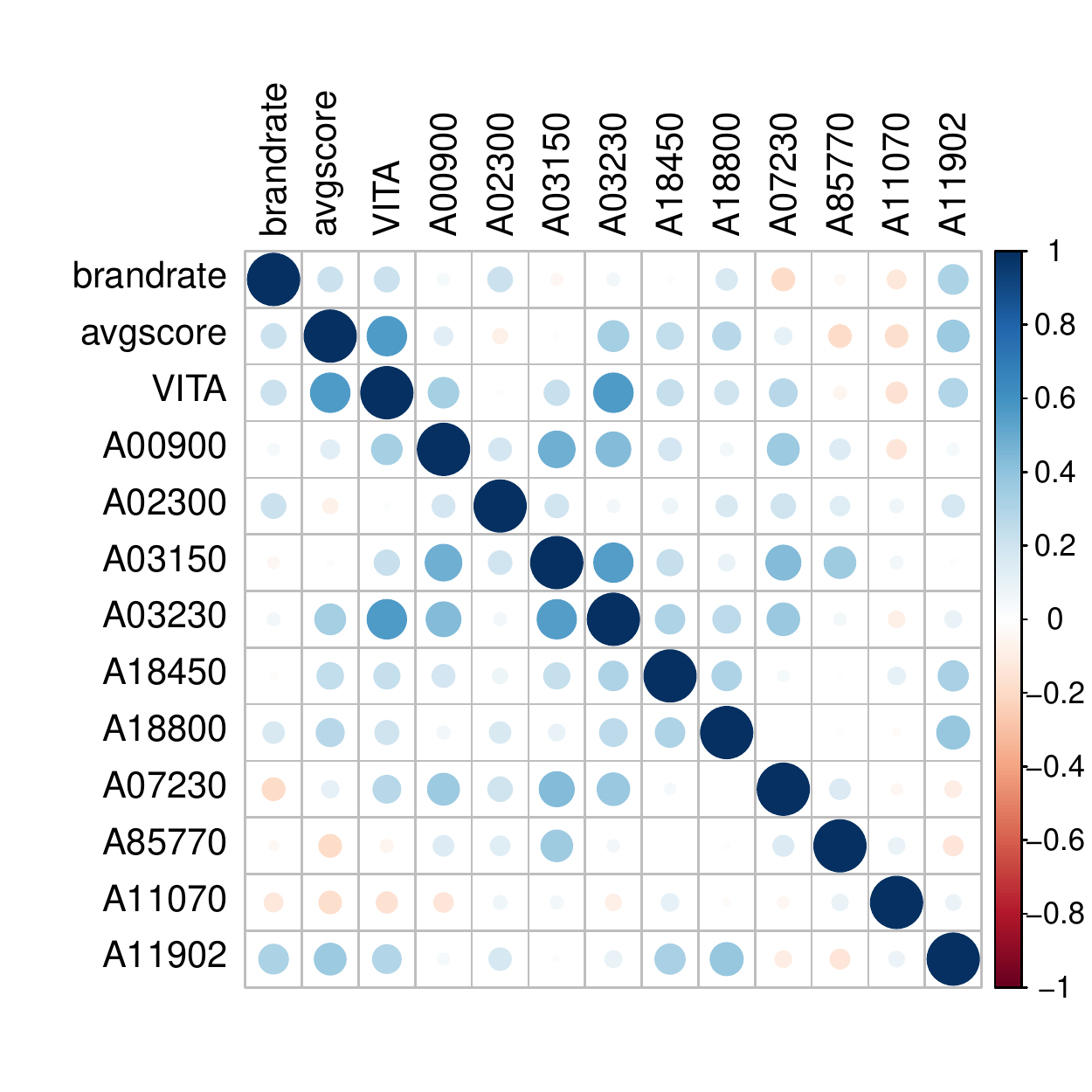}
	\caption{Correlation of the selected covariates and the response variable.}\label{fig:corr}
\end{figure}

Figure \ref{fig:corr} shows a correlation plot of the response variable and the covariates. From the figure, we see that many of these variables are positively correlated. The variable \texttt{A11070} is negatively correlated with a few other variables. 

\subsection{Fitting of the models}

Bayesian methods have been used to analyze spatial data for about twenty years since year 2000 with the advent of Markov Chain Monte Carlo (MCMC) simulative methods \citep{blangiardo2015}. In fact, the advent of MCMC has allowed researchers to use Bayesian methods to develop complex models on large datasets. However, one major drawback of MCMC methods is that they are computationally demanding, especially for large datasets. INLA (Integrated Nested Laplace Approximation) is as an alternative to MCMC  for Bayesian inference \citep{rue2005,rue2009inla}. A major advantage of INLA is that it is a deterministic algorithm and is capable of producing accurate and fast results. Since INLA was embedded into R through the R package \texttt{INLA}, it has become popular among researchers and practitioners. For example, \cite{tufvesson2019spatial} used INLA to model car insurance frequency and severity. In this paper, we use the R package \texttt{INLA} \citep{blangiardo2015,wang2018} to fit the four Beta regression models. 

The four models described in Section \ref{sec:model} can be formulated as Bayesian hierarchical models. For example, the BetaBYM model can be expressed as
\begin{subequations}
\begin{equation}
y_i\vert \eta_i \sim Beta(g^{-1}(\eta_i), \phi),
\end{equation}
\begin{equation}
\eta_i = \bfx{x}_i^T\boldsymbol{\beta} + u_i + v_i,
\end{equation}
\begin{equation}
v_i\vert\psi_1 \sim N\left(0, \psi_1^{-1}\right),
\end{equation}
\begin{equation}
\mathbf{u}\vert \psi_2 \sim N\left(\mathbf{0}, \psi_2^{-1}\mathbf{Q}^{-1}\right),
\end{equation}
\begin{equation}
\boldsymbol{\psi} \sim \pi(\boldsymbol{\psi}),
\end{equation}
\end{subequations}
where $\pi(\cdot)$ denotes a prior distribution for the two hyperparameters $\psi_1$ and $\psi_2$. Common choices for the prior distribution of $\boldsymbol{\psi}$ include independent gamma distributions. In this paper, we use the default priors in INLA.

\subsection{Model selection and validation}

The deviance information criterion (DIC) of \cite{spiegel2002dic} and the Watanabe Akaike information criterion (WAIC) of \cite{watanabe2010waic} are commonly used to select Bayesian models. The DIC is motivated by the Akaike information criterion (AIC) and is defined as
\begin{equation}\label{eq:dic}
DIC = E[D(\boldsymbol{\theta})] + p_D=D(E[\boldsymbol{\theta}]) + 2p_D,
\end{equation}
where 
\[
D(\boldsymbol{\theta}) =-2\log(p(\mathbf{y}\vert\boldsymbol{\theta}))
\]
is the deviance of the model and
$p_D$ denotes the effective number of parameters that is defined as
\[
p_D = E[D(\boldsymbol{\theta})] - D(E[\boldsymbol{\theta}]).
\]
In a Bayesian model, the deviance is random variable and the expected deviance under the posterior distribution is used as a measure of fit. Between two models, the model with a lower DIC value is preferred.

The WAIC is similar to the DIC but the effective number of parameters is calculated differently. The WAIC is defined as
\begin{equation}\label{eq:waic}
WAIC = D(E[\boldsymbol{\theta}]) + 2p_W,
\end{equation}
where 
\[
p_W = \sum_{i=1}^n \left\{ E\left[ (\log p(y_i\vert\boldsymbol{\theta}) )^2\right] - E\left[ \log p(y_i\vert\boldsymbol{\theta}) \right]^2\right\}.
\]
The WAIC is interpreted in the same way as the DIC. That is, the lower the WAIC, the more preferable the model. Between the DIC and the WAIC, \cite{gelman2014dic} argue that the WAIC is preferred. 

We also use two measures to validate the out-of-sample performance of the models. The first measure is the concordance correlation coefficient (CCC), which is used to measure the agreement between two variables. The CCC is defined as \citep{lin1989ccc}:
\begin{equation}\label{eq:ccc}
CCC = \dfrac{2 \rho \sigma_1 \sigma_2}{\sigma_1^2+\sigma_2^2 +(\mu_1-\mu_2)^2}.
\end{equation}
where $\rho$ is the correlation between the observed values and the predicted values, $\sigma_1$ and $\mu_1$ are the standard deviation and the mean of the observed values, respectively, and $\sigma_2$ and $\mu_2$ are the standard deviation and the mean of the predicted values, respectively. The value of the CCC ranges from $-1$ to 1 with a value of 1 indicating perfect agreement between the predicted values and the observed values. Between two models, a higher CCC value means a better model. 

The second measure is the well-known residual standard error (RSE), which is defined as
\begin{equation}\label{eq:rse}
RSE = \sqrt{\dfrac{1}{n} \sum_{i=1}^n (\hat{y}_i - y_i)^2},
\end{equation}
where $y_i$ and $\hat{y}_i$ represent the $i$th observed value and the $i$th predicted value, respectively. Between two models, the model with a lower RSE value is better.

\subsection{Results}

Table \ref{tbl:meai} shows the DIC and the WAIC of fitting the four Beta regression models with different link functions to the training dataset. If we look at the rows, we see that the performance based on the five link functions is quite similar except for the Cauchy link. If we look at the columns, we see that the BetaBYM model performs the best. In terms of DIC, the BetaBYM model with the Cauchy link performs the best. In terms of WAIC, however, the BetaBYM model with the log-log link is the best. However, the performance of the BetaBYM model is quite similar for the first four link functions. Using the logit link function is not a bad choice.

\begin{table}[htbp]
	\centering
	\caption{In-sample performance of the models with different link functions.}\label{tbl:meai}
\begin{tabular}{lrrrrr}
	\toprule
	\textbf{Model} & \textbf{Logit} & \textbf{Probit} & \textbf{Loglog} & \textbf{Cloglog} & \textbf{Cauchy} \\
	\midrule
	& \multicolumn{5}{c}{$DIC$} \\
	BetaReg & -2103.82 & -2102.92 & -2101.53 & -2104.36 & -2109.25 \\
	BetaRE & -2103.60 & -2102.20 & -2100.60 & -2104.10 & -2109.79 \\
	BetaBesag & -2105.79 & -2109.89 & -2109.58 & -2107.41 & -2114.87 \\
	BetaBYM & -2116.93 & -2115.13 & -2112.55 & -2118.06 & \textbf{-2132.92} \\
	\midrule
	& \multicolumn{5}{c}{$WAIC$} \\
	BetaReg & -2070.47 & -2070.84 & -2071.36 & -2069.72 & -2062.49 \\
	BetaRE & -2070.60 & -2071.07 & -2071.56 & -2069.88 & -2063.02 \\
	BetaBesag & -2077.25 & -2081.93 & -2083.42 & -2077.41 & -2069.28 \\
	BetaBYM & -2089.08 & -2091.05 & \textbf{-2091.60} & -2088.89 & -2087.56 \\
	\bottomrule
\end{tabular}%
\end{table}%

\begin{table}[htbp]
	\centering
	\caption{Out-of-sample performance of the models with different link functions.}\label{tbl:meao}
\begin{tabular}{lrrrrr}
	\toprule
	\textbf{Model} & \textbf{Logit} & \textbf{Probit} & \textbf{Loglog} & \textbf{Cloglog} & \textbf{Cauchy} \\
	\midrule
	& \multicolumn{5}{c}{$CCC$} \\
	BetaReg & 0.42846 & 0.42810 & 0.42749 & 0.42801 & 0.42138 \\
	BetaRE & 0.42834 & 0.42790 & 0.42733 & 0.42786 & 0.42126 \\
	BetaBesag & 0.45289 & 0.46855 & 0.47168 & 0.45622 & 0.44610 \\
	BetaBYM & 0.48595 & 0.48954 & \textbf{0.49127} & 0.48584 & 0.47814 \\
	\midrule
	& \multicolumn{5}{c}{$RSE$} \\
	BetaReg & 0.01380 & 0.01376 & 0.01371 & 0.01383 & 0.01412 \\
	BetaRE & 0.01380 & 0.01376 & 0.01371 & 0.01383 & 0.01413 \\
	BetaBesag & 0.01364 & 0.01352 & 0.01342 & 0.01367 & 0.01412 \\
	BetaBYM & 0.01359 & 0.01352 & \textbf{0.01340} & 0.01367 & 0.01445 \\
	\bottomrule
\end{tabular}%
\end{table}%

Table \ref{tbl:meao} show the out-of-sample performance of the models with different link functions on the test set. In terms of the CCC, the BetaBYM model with the log-log link performs the best because the corresponding CCC value is the highest among the 20 cases. In terms of the RSE, the BetaBYM model with the log-log link is also the best as it has the lowest RSE value.

\begin{figure}[htbp]
	\centering
	\begin{subfigure}{0.49\textwidth}
		\includegraphics[width=\textwidth]{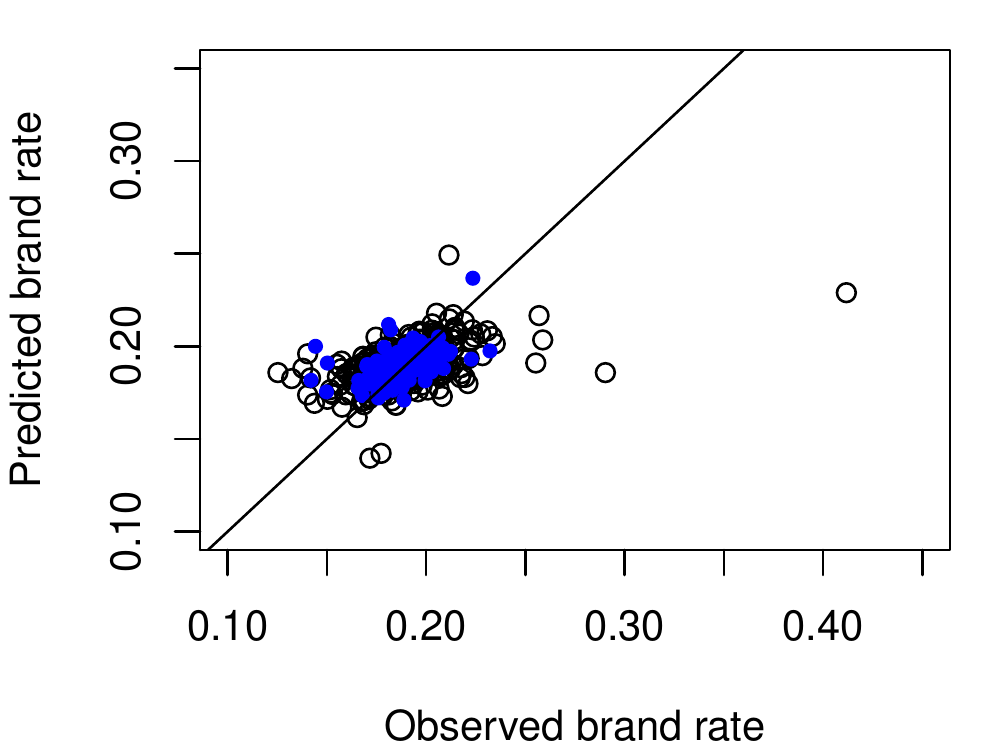}
		\caption{The basic Beta regression model.}
	\end{subfigure}
	\begin{subfigure}{0.49\textwidth}
		\includegraphics[width=\textwidth]{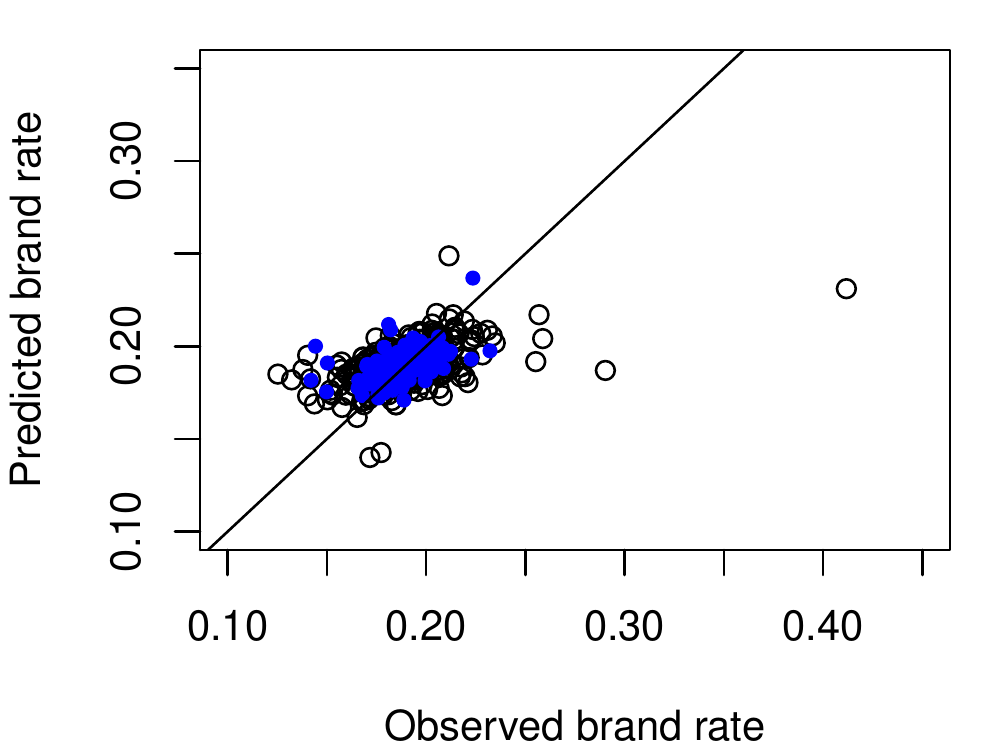}
		\caption{The BetaRE model.}
	\end{subfigure}
\begin{subfigure}{0.49\textwidth}
	\includegraphics[width=\textwidth]{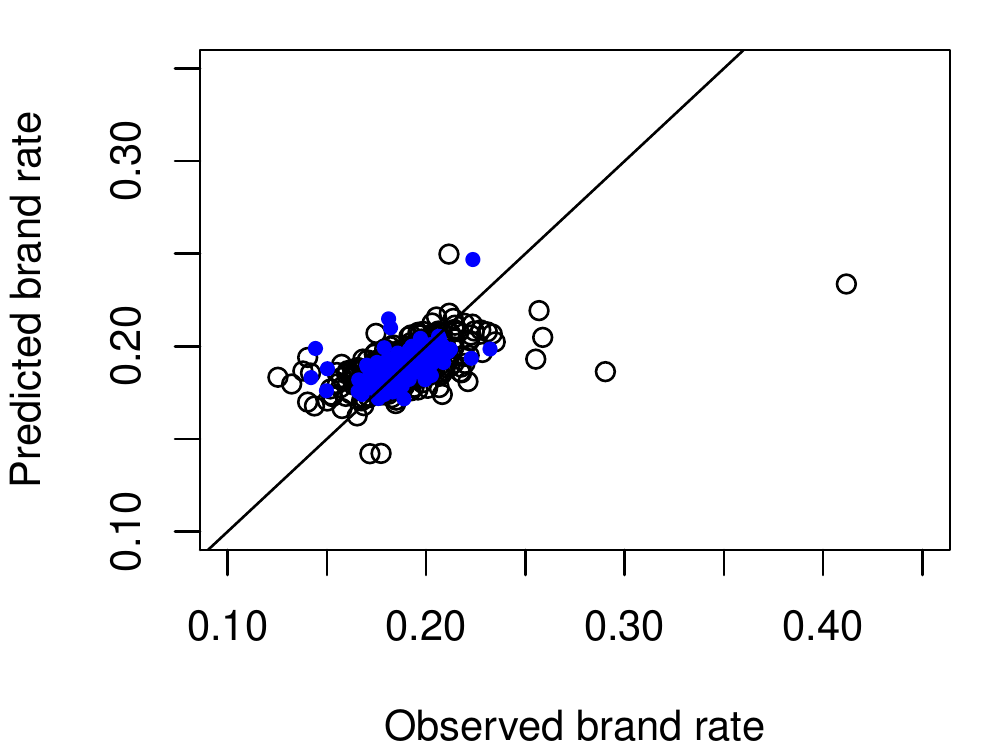}
	\caption{The BetaBesag model.}
\end{subfigure}
\begin{subfigure}{0.49\textwidth}
	\includegraphics[width=\textwidth]{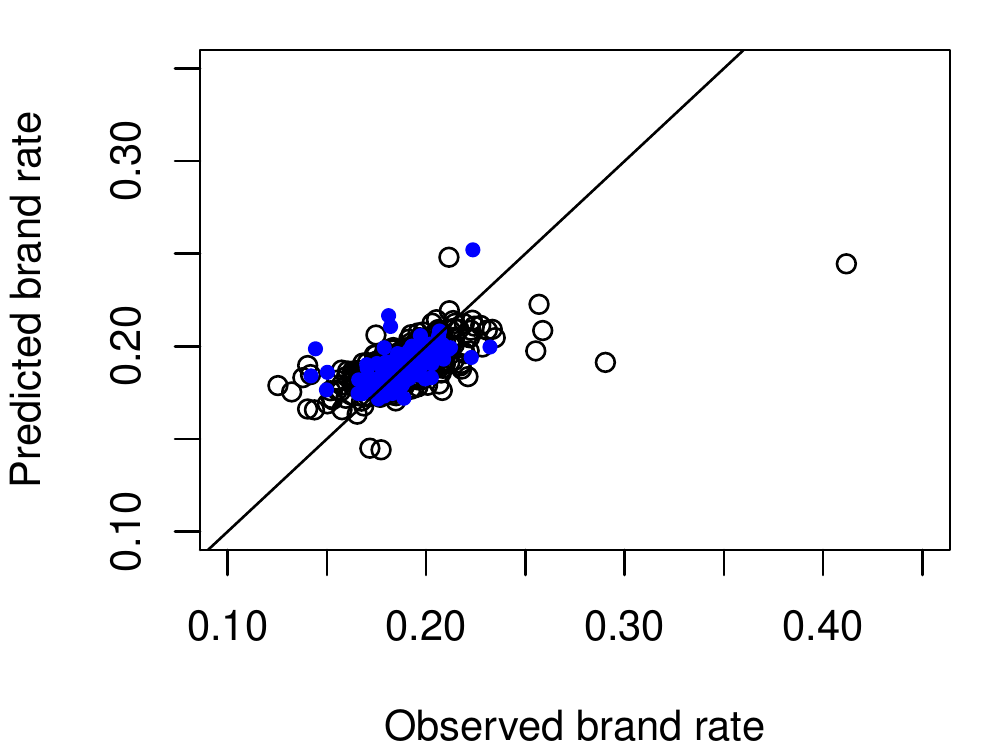}
	\caption{The BetaBYM model.}
\end{subfigure}
\caption{Scatter plots of the observed brand name drug claim rates and  those predicted by the four models with the log-log link function. The out-of-sample predictions are plotted as  blue dots, while the in-sample fitted values are plotted as black circles.}\label{fig:betascatter}
\end{figure}

\begin{figure}[htbp]
	\centering
	\begin{subfigure}{0.49\textwidth}
		\includegraphics[width=\textwidth]{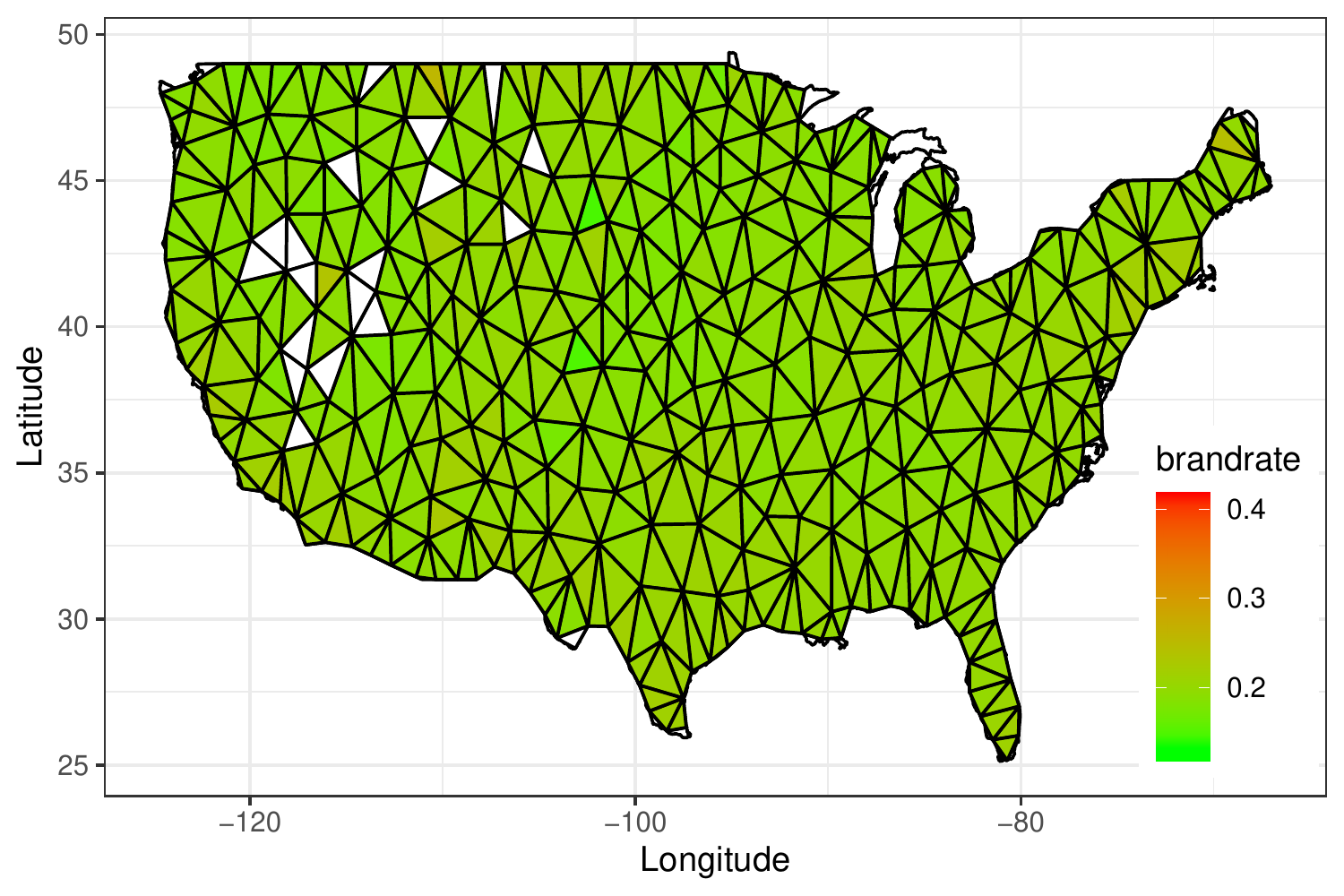}
		\caption{The basic Beta regression model.}
	\end{subfigure}
	\begin{subfigure}{0.49\textwidth}
		\includegraphics[width=\textwidth]{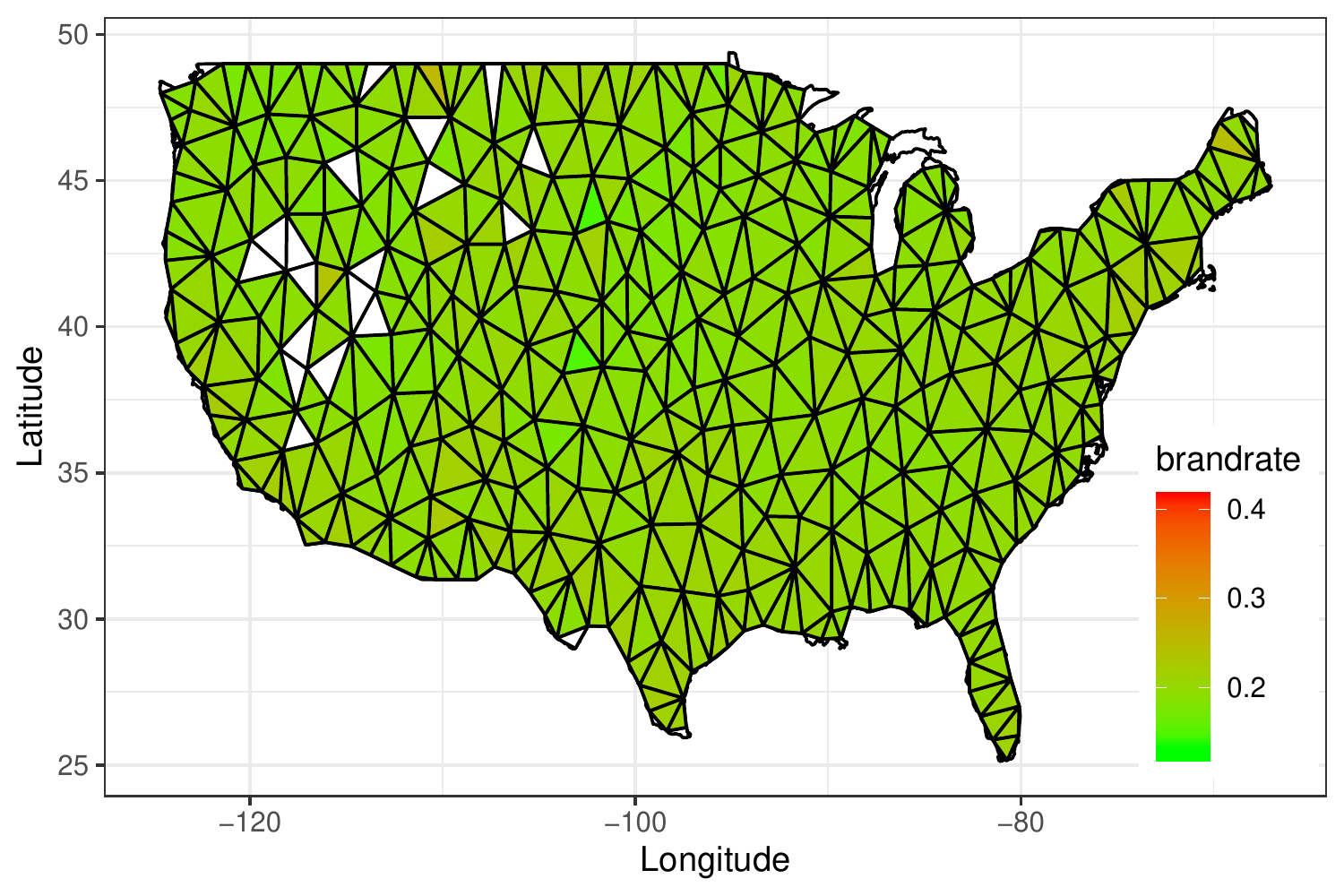}
		\caption{The BetaRE model.}
	\end{subfigure}
	\begin{subfigure}{0.49\textwidth}
		\includegraphics[width=\textwidth]{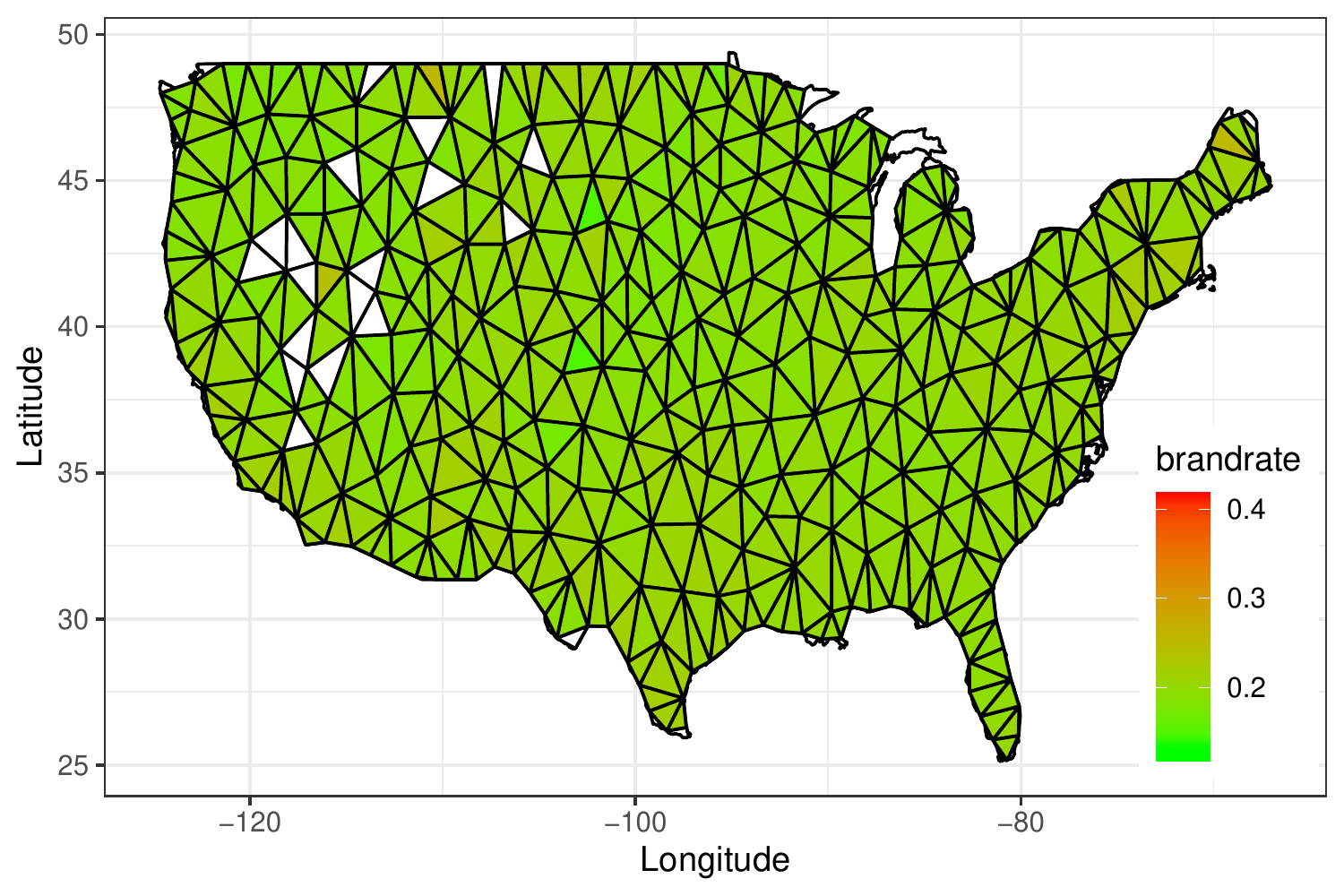}
		\caption{The BetaBesag model.}
	\end{subfigure}
	\begin{subfigure}{0.49\textwidth}
		\includegraphics[width=\textwidth]{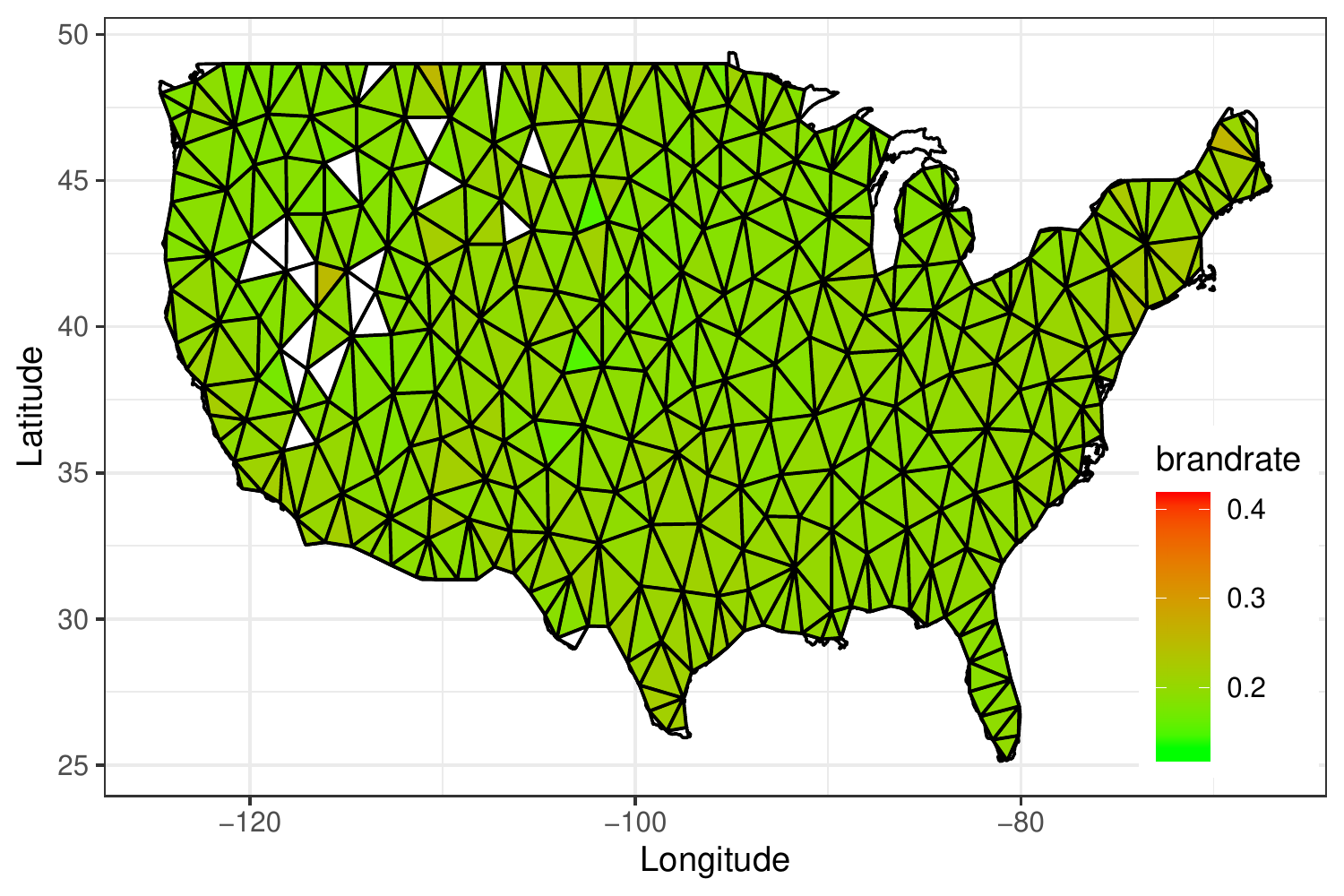}
		\caption{The BetaBYM model.}
	\end{subfigure}
	\caption{Distributions of the brand name drug claim rates predicted by the four models with the log-log link function.}\label{fig:betam}
\end{figure}

Figure \ref{fig:betascatter} shows the scatter plots of the observed brand name drug claim rates and the predicted values produced by the four models with the log-log link function. From the four scatter plots, we see that the predicted values of the four models are pretty similar. We also see  that all four models do not fit the large rates well. The highest predicted value is around 0.25, while the largest observed value is around 0.41 (see Table \ref{tbl:data1}). Figure \ref{fig:betam} shows the distributions of the brand name drug claim rates predicted by the four models with the log-log link function across the triangular areas. The maps look quite similar and it is hard to see the differences. In addition, these maps look similar to the map of the observed rates shown in Figure \ref{fig:rate}.

\begin{table}[htbp]
  \centering
  \caption{Coefficients and hyperparameters from fitting the basic Beta regression model with the log-log link.}\label{tbl:betareg}
    \begin{tabular}{lrrrrr}
    \toprule
          & \textbf{Mean} & \textbf{Std} & \textbf{0.025 Q} & \textbf{0.5 Q} & \textbf{0.975 Q.} \\
          \midrule
    (Intercept) & -0.554 & 0.106 & -0.763 & -0.554 & -0.345 \\
    \texttt{avgscore} & 0.061 & 0.027 & 0.009 & 0.061 & 0.113 \\
    \texttt{VITA} & 0.003 & 0.001 & 0.001 & 0.003 & 0.006 \\
    \texttt{A00900} & 0.004 & 0.010 & -0.017 & 0.004 & 0.024 \\
    \texttt{A02300} & 0.066 & 0.014 & 0.039 & 0.066 & 0.093 \\
    \texttt{A03150} & 0.003 & 0.011 & -0.018 & 0.003 & 0.025 \\
    \texttt{A03230} & -0.002 & 0.013 & -0.027 & -0.002 & 0.023 \\
    \texttt{A18450} & -0.044 & 0.013 & -0.069 & -0.044 & -0.019 \\
    \texttt{A18800} & 0.002 & 0.029 & -0.055 & 0.002 & 0.060 \\
    \texttt{A07230} & -0.198 & 0.044 & -0.285 & -0.198 & -0.111 \\
    \texttt{A85770} & 0.012 & 0.011 & -0.010 & 0.012 & 0.035 \\
    \texttt{A11070} & -0.234 & 0.103 & -0.437 & -0.234 & -0.031 \\
    \texttt{A11902} & 0.132 & 0.039 & 0.057 & 0.133 & 0.208 \\
    $\phi$ & 388.740 & 27.220 & 337.310 & 388.040 & 444.170 \\
    \bottomrule
    \end{tabular}%
\end{table}%

\begin{table}[htbp]
  \centering
  \caption{Coefficients and hyperparameters from fitting the Beta regression model with random effects. The log-log link function was used.}\label{tbl:betare}
    \begin{tabular}{lrrrrr}
    \toprule
          & \textbf{Mean} & \textbf{Std} & \textbf{0.025 Q} & \textbf{0.5 Q} & \textbf{0.975 Q.} \\
          \midrule
    (Intercept) & -0.553 & 0.107 & -0.763 & -0.553 & -0.344 \\
    \texttt{avgscore} & 0.061 & 0.027 & 0.008 & 0.061 & 0.113 \\
    \texttt{VITA} & 0.003 & 0.001 & 0.001 & 0.003 & 0.006 \\
    \texttt{A00900} & 0.004 & 0.010 & -0.017 & 0.004 & 0.024 \\
    \texttt{A02300} & 0.066 & 0.014 & 0.039 & 0.066 & 0.093 \\
    \texttt{A03150} & 0.003 & 0.011 & -0.018 & 0.003 & 0.025 \\
    \texttt{A03230} & -0.002 & 0.013 & -0.027 & -0.002 & 0.023 \\
    \texttt{A18450} & -0.044 & 0.013 & -0.069 & -0.044 & -0.019 \\
    \texttt{A18800} & 0.002 & 0.029 & -0.055 & 0.002 & 0.060 \\
    \texttt{A07230} & -0.198 & 0.045 & -0.285 & -0.198 & -0.111 \\
    \texttt{A85770} & 0.012 & 0.011 & -0.010 & 0.012 & 0.035 \\
    \texttt{A11070} & -0.234 & 0.104 & -0.438 & -0.234 & -0.030 \\
    \texttt{A11902} & 0.133 & 0.039 & 0.057 & 0.133 & 0.208 \\
    $\phi$ & 390.650 & 27.960 & 337.970 & 389.880 & 448.040 \\
    $\psi_1$ & 30149.160 & 22026.050 & 5634.800 & 24634.420 & 87681.270 \\
    \bottomrule
    \end{tabular}%
\end{table}%

\begin{table}[htbp]
  \centering
  \caption{Coefficients and hyperparameters from fitting the BetaBesag  model  with the log-log link.}\label{tbl:betabesag}
    \begin{tabular}{lrrrrr}
    \toprule
          & \textbf{Mean} & \textbf{Std} & \textbf{0.025 Q} & \textbf{0.5 Q} & \textbf{0.975 Q.} \\
          \midrule
(Intercept) & -0.574 & 0.109 & -0.788 & -0.574 & -0.361 \\
\texttt{avgscore} & 0.067 & 0.027 & 0.013 & 0.067 & 0.121 \\
\texttt{VITA} & 0.003 & 0.001 & 0.001 & 0.003 & 0.006 \\
\texttt{A00900} & 0.002 & 0.011 & -0.018 & 0.002 & 0.023 \\
\texttt{A02300} & 0.065 & 0.014 & 0.037 & 0.065 & 0.093 \\
\texttt{A03150} & 0.004 & 0.011 & -0.017 & 0.004 & 0.026 \\
\texttt{A03230} & 0.000 & 0.013 & -0.025 & 0.000 & 0.025 \\
\texttt{A18450} & -0.036 & 0.014 & -0.064 & -0.036 & -0.007 \\
\texttt{A18800} & -0.012 & 0.031 & -0.074 & -0.012 & 0.049 \\
\texttt{A07230} & -0.215 & 0.048 & -0.310 & -0.215 & -0.121 \\
\texttt{A85770} & 0.012 & 0.011 & -0.011 & 0.012 & 0.034 \\
\texttt{A11070} & -0.206 & 0.110 & -0.421 & -0.206 & 0.011 \\
\texttt{A11902} & 0.134 & 0.039 & 0.057 & 0.134 & 0.210 \\
$\phi$ & 411.830 & 31.980 & 350.770 & 411.430 & 476.320 \\
$\psi_2$ & 3483.030 & 5884.000 & 461.760 & 1839.520 & 16609.730 \\
\bottomrule
    \end{tabular}%
\end{table}%

\begin{table}[htbp]
  \centering
  \caption{Coefficients and hyperparameters from fitting the BetaBYM  model with the log-log link.}\label{tbl:betabym}
    \begin{tabular}{lrrrrr}
    \toprule
          & \textbf{Mean} & \textbf{Std} & \textbf{0.025 Q} & \textbf{0.5 Q} & \textbf{0.975 Q.} \\
          \midrule
(Intercept) & -0.580 & 0.111 & -0.800 & -0.580 & -0.362 \\
\texttt{avgscore} & 0.071 & 0.028 & 0.016 & 0.071 & 0.126 \\
\texttt{VITA} & 0.003 & 0.001 & 0.001 & 0.003 & 0.006 \\
\texttt{A00900} & 0.001 & 0.011 & -0.020 & 0.001 & 0.023 \\
\texttt{A02300} & 0.065 & 0.015 & 0.036 & 0.065 & 0.093 \\
\texttt{A03150} & 0.005 & 0.011 & -0.017 & 0.005 & 0.026 \\
\texttt{A03230} & 0.002 & 0.013 & -0.024 & 0.002 & 0.028 \\
\texttt{A18450} & -0.031 & 0.014 & -0.059 & -0.031 & -0.002 \\
\texttt{A18800} & -0.021 & 0.031 & -0.083 & -0.021 & 0.041 \\
\texttt{A07230} & -0.225 & 0.049 & -0.321 & -0.225 & -0.128 \\
\texttt{A85770} & 0.011 & 0.012 & -0.012 & 0.011 & 0.034 \\
\texttt{A11070} & -0.195 & 0.114 & -0.419 & -0.196 & 0.028 \\
\texttt{A11902} & 0.135 & 0.040 & 0.056 & 0.135 & 0.213 \\
$\phi$ & 425.630 & 33.890 & 362.380 & 424.470 & 495.770 \\
$\psi_1$ & 6583.970 & 2917.140 & 2614.040 & 6021.720 & 13846.680 \\
$\psi_2$ & 1435.850 & 1101.810 & 384.650 & 1118.950 & 4360.820 \\
\bottomrule
    \end{tabular}%
\end{table}%

Tables \ref{tbl:betareg}, \ref{tbl:betare}, \ref{tbl:betabesag}, and \ref{tbl:betabym} show the fitted coefficients and hyperparameters of the four models with the log-log link function. In particular, the tables show the mean, the standard deviation, the 2.5\% quantile, the median, and the 97.5\% quantile of the estimated parameters. From the 2.5\% and the 97.5\% quantiles, we get the 95\% credibility intervals of the fitted parameters. If we look at the credibility intervals of the fitted coefficients, we see that the credibility intervals of \texttt{avgscore}, \texttt{VITA}, \texttt{A02300}, and \texttt{A11902} contain only positive values. This means that these variables tend to have positive impact on the responses, i.e., the brand name drug claim rates. 

From these tables, we also see two variables whose 95\% credibility intervals contain only negative values. The two variables are \texttt{A07230} and \texttt{A18450}. From Table \ref{tbl:covariate}, we see that both variables are related to education spending. Since the fitted coefficients of these two variables are negative, the two variables tend to have negative impact on the brand name drug claim rates.

Tables \ref{tbl:betare}, \ref{tbl:betabesag}, and \ref{tbl:betabym} also show the estimated hyperparameters of the unstructured random effects and the structured random effects. Note that the hyperparameters are precision parameters that are inverse to the variance of the random effects. The higher the hyperparameter, the lower the variance of the random effects. In Table \ref{tbl:betare}, we see that the estimated hyperparameter $\psi_1$ is very large. This means that the random effects component of the model has a very small variance and does not contribute much to explain the variation of the response variable. 

In Table \ref{tbl:betabesag}, we see that the fitted precision parameter $\psi_2$ of the structured random effects is much lower than the fitted $\psi_1$ in Table \ref{tbl:betare}. In Table \ref{tbl:betabym}, we see the same pattern that fitted $\psi_2$ is much lower than the fitted $\psi_1$. This means that the structured random effects help more than the unstructured random effects in terms of explaining the response variation. 

In summary, the above results show that modeling the spatial effect improves the performance of the Beta regression model. In addition, including unstructured random effects in the model only improve the performance marginally. To further assess the comprehensiveness of our Beta regression models, we provide additional results using models suggested in \cite{duncan2016health}. See Appendix \ref{sec:addres} for additional results.

\section{Concluding Remarks}\label{sec:con}

The healthcare sector in the U.S. is a large sector that generates about 20\% of the country's gross domestic product. This significance of healthcare has motivated researchers to develop enhanced tools and approaches to better understand the industry through healthcare analytics. According to the 2019 Predictive Analytics in Health Care Trend Forecast \citep{soa2019survey}, predictive analytics is poised to reshape the healthcare industry by achieving three aims: improved patient outcomes, higher quality of care, and reduced costs. In particular,  McKinsey estimates that big data analytics can help the U.S. healthcare industry to save more than \$300 billion per year, where two thirds of that come from the reductions of approximately 8\% in national healthcare expenditures.

In this paper, we examined and demonstrated the use of Beta regression models to study the utilization of brand name drugs in the U.S. in order to understand variability of the brand name drug claim rates across different areas of the U.S.. We studied different Beta regression models with and without spatial effects and fitted these models to public datasets obtained from the CMS and the IRS. The numerical results show that Beta regression models are able to fit the brand name drug claim rates well and modeling the spatial effects improves the performance of the Beta regression models. We also find some significant variables that help to explain the variation of the brand name drug claim rates across different areas. The methods and findings in this paper are useful for healthcare actuaries in their data analysis. By reflecting the effect of prescription drug utilization, these models can be used to update an insured's risk score in a risk adjustment model. Specifically, healthcare actuaries can incorporate the geographic variation in their models used to select preferred providers.

\bibliographystyle{apalike}
\bibliography{hcdrug}

\appendix

\section{Additional results}\label{sec:addres}

\cite{duncan2016health} tested several alternative regression frameworks for predictive modeling of health care costs. In particular, they tested multiple linear regression models, lasso, multivariate adaptive regression splines (MARS), random forests, M5 decision trees, and boosted trees. For details of these models, readers are referred to \cite{duncan2016health} and the references therein. In this section, we present the out-of-sample prediction results of these models.

\begin{table}[htbp]
  \centering
  \caption{Parameters and the R packages of the additional models.}\label{tbl:addmodel}
    \begin{tabular}{lll}
    \toprule
    \textbf{Model} & \textbf{R Package} & \textbf{Parameters} \\
    \midrule
    Linear model & \texttt{lm} & Multiple linear regression without interactions \\
    Lasso & \texttt{glmnet} & Gaussian family \\
    MARS  & \texttt{earth} & Interactions up to degree = 1; \\
    & &Generalized Cross Validation penalty  = 1 \\
    Random forests & \texttt{randomForest} & Number of trees = 100; minimum node size = 10 \\
    M5 decision trees & \texttt{Cubist} & Number of committees = 25 \\
    Boosted trees & \texttt{gbm} & Number of trees = 5000; Shrinkage = 0.01 \\
    \bottomrule
    \end{tabular}%
\end{table}%

\begin{table}[htbp]
  \centering
  \caption{Out-of-sample prediction results of the additional models.}\label{tbl:addresult}
    \begin{tabular}{lll}
    \toprule
    \textbf{Model} & \boldmath{}\textbf{$CCC$}\unboldmath{} & \boldmath{}\textbf{$RSE$}\unboldmath{} \\
    \midrule
    Linear model & 0.4129 & 0.0141 \\
    Lasso & 0.0000 & 0.0154 \\
    MARS  & 0.4420 & 0.0143 \\
    Random forest & 0.4657 & 0.0126 \\
    M5 decision tree & 0.4330 & 0.0131 \\
    Boosted trees & 0.4658 & 0.0133 \\
    \bottomrule
    \end{tabular}%
\end{table}%

\begin{figure}[htbp]
	\centering
	\begin{subfigure}{0.49\textwidth}
		\includegraphics[width=\textwidth]{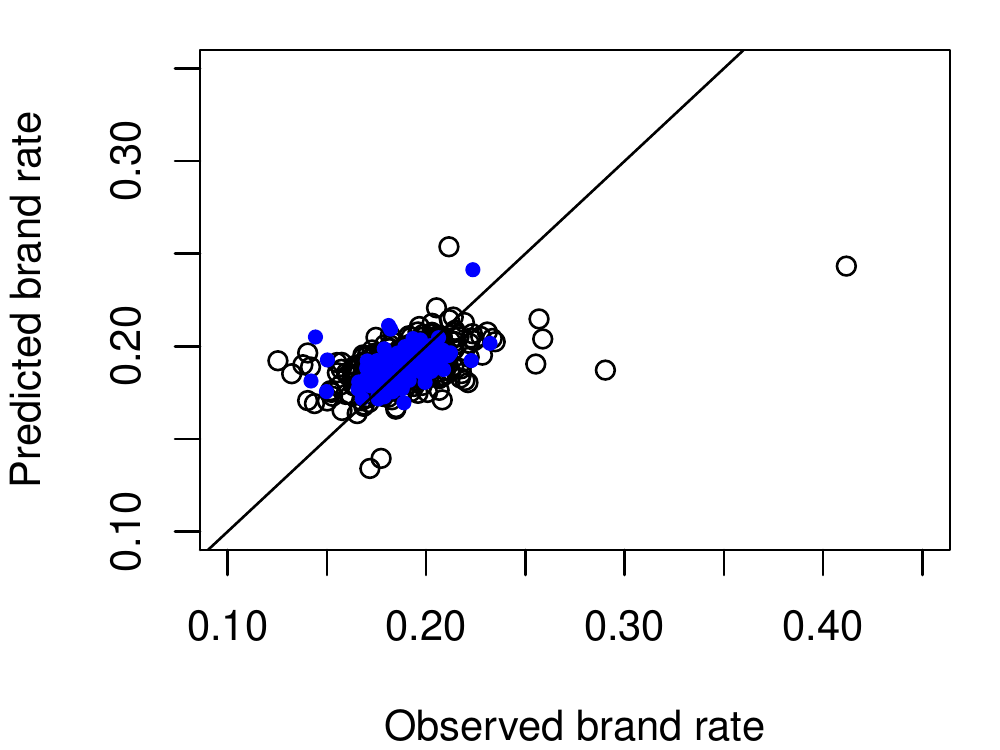}
		\caption{The linear regression model.}
	\end{subfigure}
	\begin{subfigure}{0.49\textwidth}
		\includegraphics[width=\textwidth]{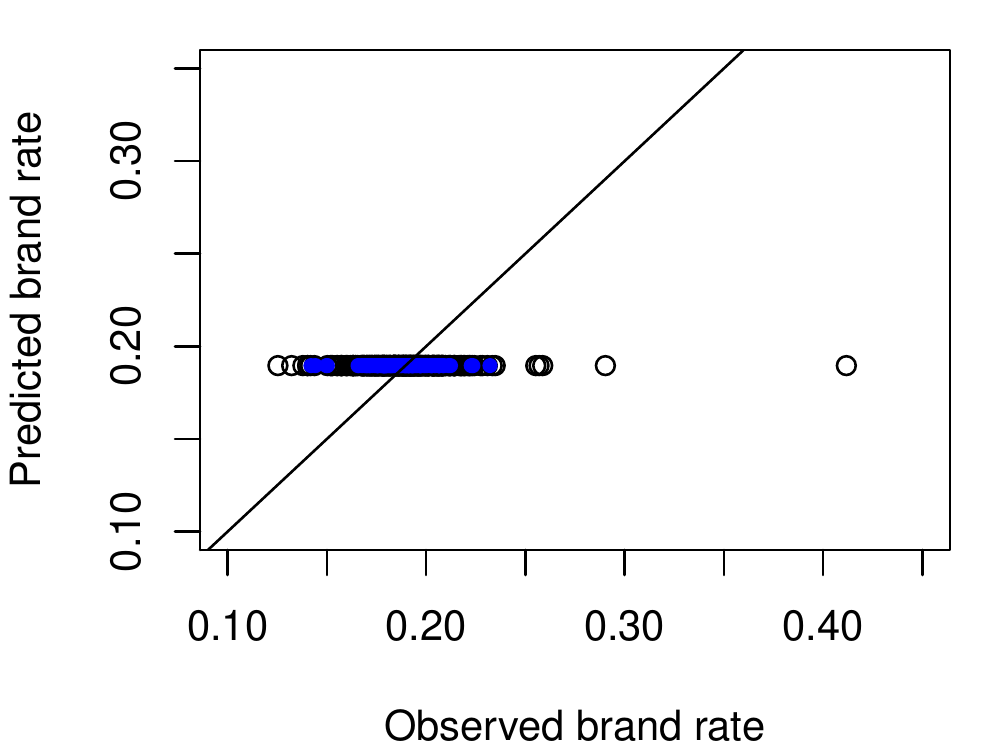}
		\caption{The lasso model.}\label{fig:lasso}
	\end{subfigure}
	\begin{subfigure}{0.49\textwidth}
		\includegraphics[width=\textwidth]{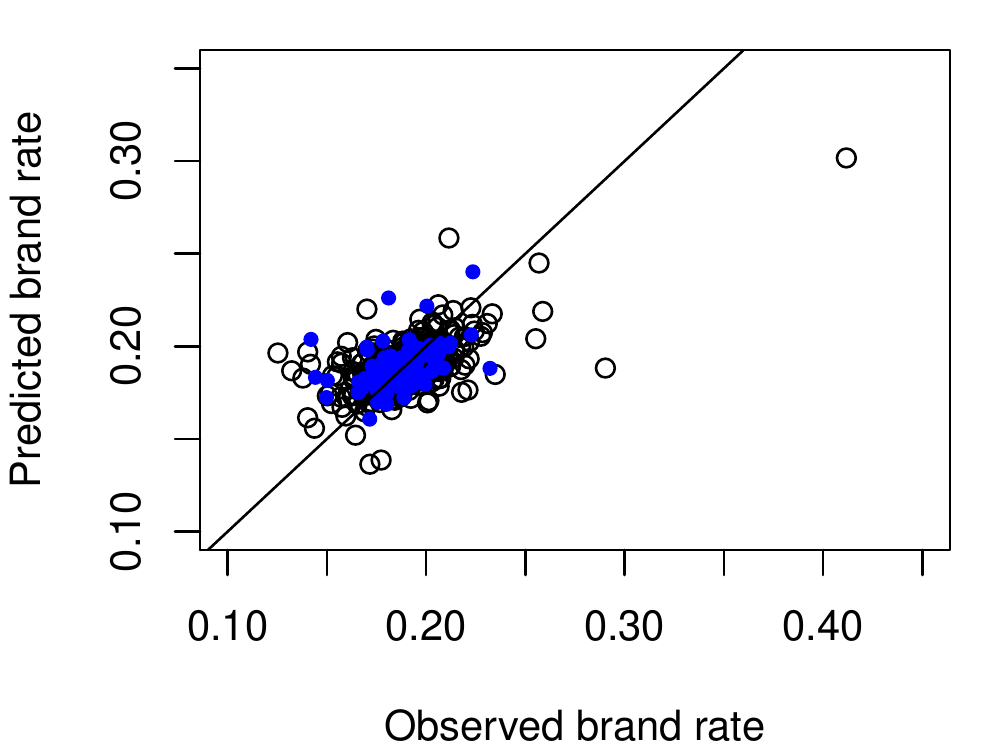}
		\caption{The MARS model.}
	\end{subfigure}
	\begin{subfigure}{0.49\textwidth}
		\includegraphics[width=\textwidth]{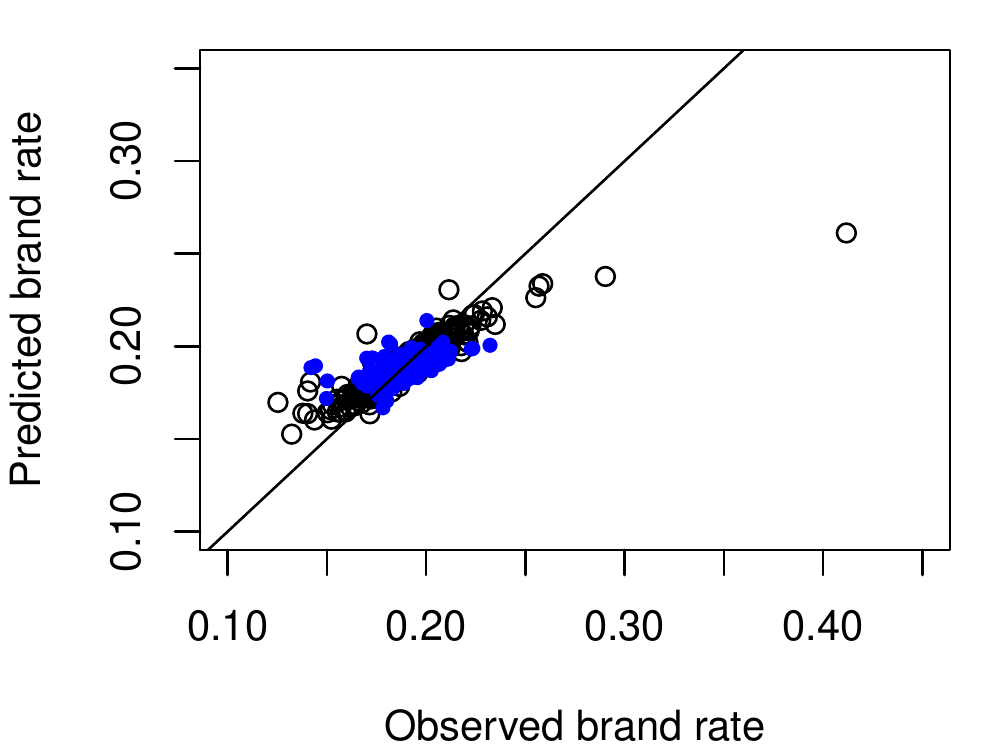}
		\caption{The random forests model.}
	\end{subfigure}
    \begin{subfigure}{0.49\textwidth}
    		\includegraphics[width=\textwidth]{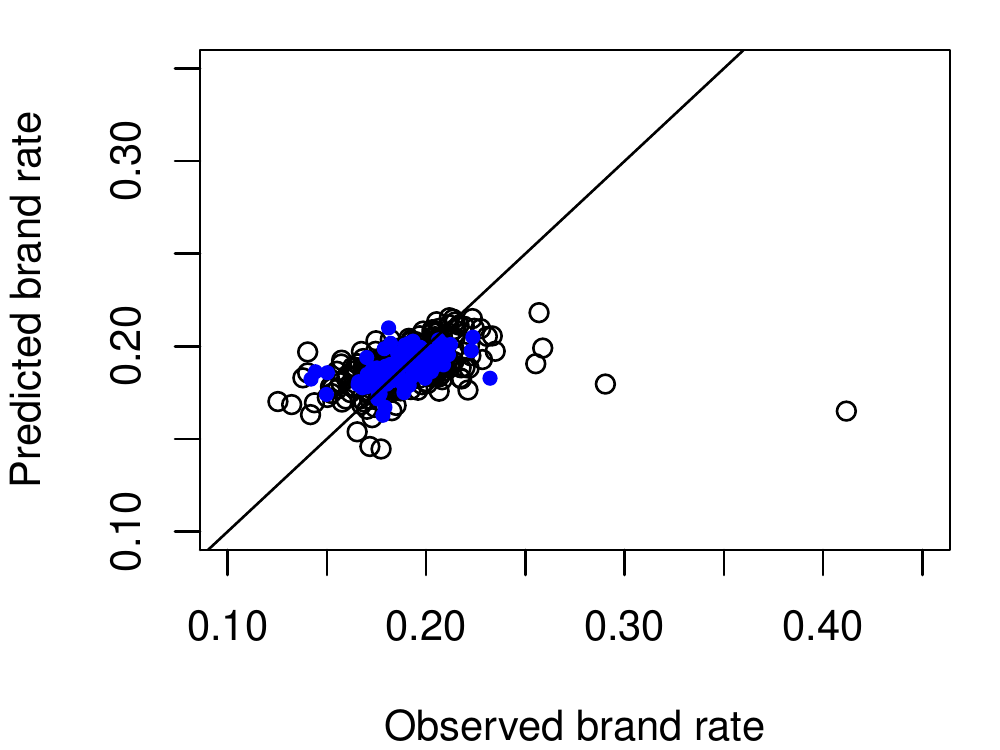}
    		\caption{The M5 decision trees model.}
    	\end{subfigure}
        \begin{subfigure}{0.49\textwidth}
        		\includegraphics[width=\textwidth]{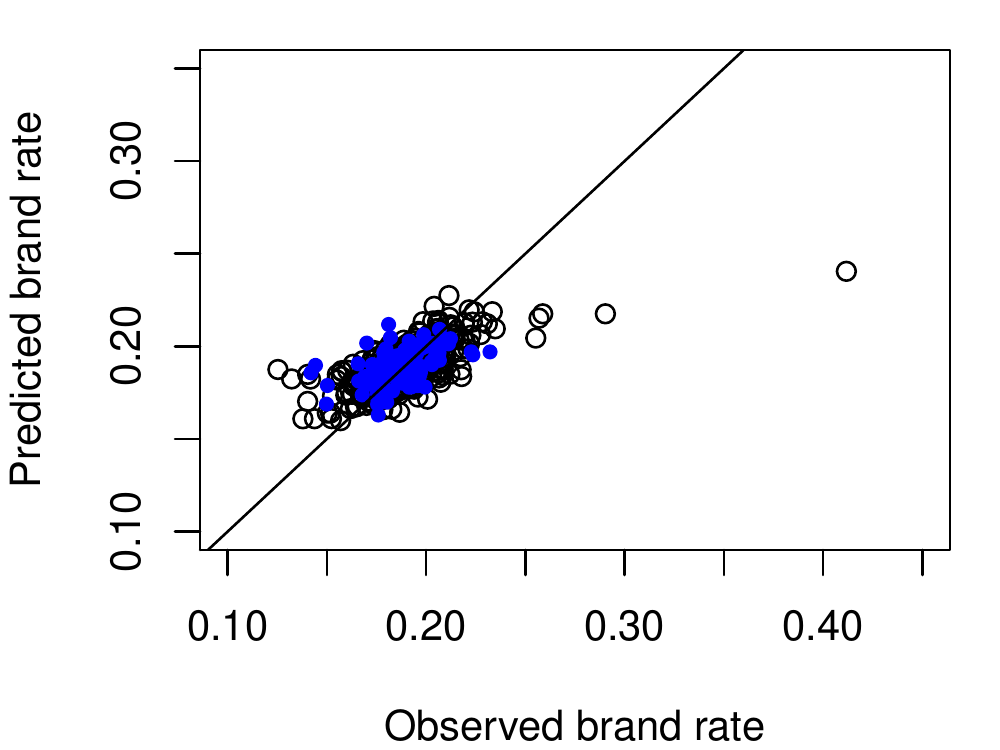}
        		\caption{The boosted trees model.}
        	\end{subfigure}
	\caption{Scatter plots of the observed brand name drug claim rates and those predicted by the additional models.}\label{fig:addresult}
\end{figure}

Table \ref{tbl:addmodel} shows the parameters and R packages we used to test the additional models. Table \ref{tbl:addresult} shows the out-of-sample prediction results obtained by these models. Figure \ref{fig:addresult} shows the scatter plots of the observed brand name drug claim rates and the predicted values by the additional models.  The variables input to these models are the same as those selected for the Beta regression models described in Table \ref{tbl:covariate}. 

From Table \ref{tbl:addresult}, we see that the lasso model produced the worst result. The reason is that the lasso regression model produced an intercept-only model for the data. Figure \ref{fig:lasso} also confirms this as the predicted values are constant. Comparing Tables \ref{tbl:meao} and \ref{tbl:addresult}, we see that all these models do not perform better than the Beta regression model in terms of $CCC$. However, the tree-based models (i.e., random forests, M5 decision trees, boosted trees) produced lower $RSE$ than the Beta regression model.

\end{document}